# MOSS II: mid-frequency radio catalogue of the Saraswati core region

R. Kincaid ,[1★] E. Retana-Montenegro,[2,3] B. Šlaus,[4,5] V. Parekh,[6] P. Jablonka,[1] S. Salunkhe,[7] S. Sankhyayan ,[8] V. Smolčić[5] and M. Bondi[9]

[1]*Institute of Physics, École Polytechnique Fédérale de Lausanne (EPFL), CH-1015 Lausanne, Switzerland*
[2]*Astrophysics Research Centre, School of Mathematics, Statistics and Computer Science, University of KwaZulu-Natal, Durban 4041, South Africa*
[3]*Wits Centre for Astrophysics, School of Physics, University of the Witwatersrand, Private Bag 3, Johannesburg 2050, South Africa*
[4]*Astronomical Observatory Institute, Faculty of Physics, Adam Mickiewicz University, ul. Słoneczna 36, PL-60-286 Poznań, Poland*
[5]*Department of Physics, Faculty of Science, University of Zagreb, Bijenička cesta 32, 10000 Zagreb, Croatia*
[6]*National Radio Astronomy Observatory (NRAO), 1011 Lopezville Rd, Socorro, NM 87801, USA*
[7]*National Centre for Radio Astrophysics, Tata Institute of Fundamental Research, Pune 411007, India*
[8]*Tartu Observatory, University of Tartu, Observatooriumi 1, 61602 Tõravere, Estonia*
[9]*INAF Institute of Astrophysics and Space Science, Via P. Gobetti 93/3, I-40129 Bologna, Italy*

## ABSTRACT

The MeerKAT Observations of the *Saraswati* Supercluster is an ongoing project attempting to study the radio and optical properties of the core region of the *Saraswati* supercluster which will eventually entail a full survey of the entire supercluster region. We have used MeerKAT *L*-band (1.28 GHz) images at an angular resolution of 8 arcsec from previous deep (central RMS noise of 11-16 μJy beam$^{-1}$) pilot observations of the core region ($z \sim 0.28$) of the *Saraswati* supercluster containing the two most massive galaxy clusters: Abell 2631 and ZwCl2341. These cluster fields cover an area of $\sim$1.6 deg$^2$ and the radio catalogues produced from each cluster region contain 1999 and 2611 sources (5$\sigma$ limit) for Abell 2631 and ZwCL2341, respectively. For each catalogue, we investigated the noise properties, astrometry, flux density scale accuracy, spectral properties, etc. of the radio sources. The catalogues were then corrected for various observational biases before derivation of the radio source counts. In agreement with previous studies, we find that at the sub-mJy level our counts show the characteristic flattening, indicating the increased dominance of the star-forming galaxy (SFG) population over the active galactic nuclei (AGN). Furthermore, in this sub-mJy regime the counts lie slightly higher (a 'bump' feature) compared to other deep MeerKAT data and recent radio-sky simulations. We suggest that this feature could be attributed to an enhanced population of intermediate SFG and/or AGNs associated with these galaxy cluster fields. In addition cosmic variance could represent an important source of uncertainty in the source counts.

## 1 INTRODUCTION

Radio continuum surveys have long played a major role in advancing observational cosmology and galactic astronomy by addressing astrophysical properties and cosmological evolution of radio galaxies, quasars, and star-forming galaxies (E. Fomalont 1999). In recent years, radio interferometers such as the Karl G. Jansky Very Large Array (VLA), LOw Frequency ARay (LOFAR), and Giant Metrewave Telescope(GMRT) have surveyed fields of different sizes (few arcmins to thousands of square degrees), depths (sub-mJy to Jy), and resolution (sub-arcseconds to arcmin) to produce several Large-area radio surveys such as the NRAO VLA Sky Survey, Faint Images of the Radio Sky at 20 cm (FIRST; R. L. White et al. 1997), Very Large Array all Sky Survey (VLASS; M. Lacy et al. 2020), LOFAR Two meter Sky Survey (T. Shimwell et al. 2017), and TIFR GMRT Sky Survey (H. Intema et al. 2017). These surveys have given us a better understanding of the universe in the radio regime. Radio observations offer an unobstructed dust-unbiased view of star formation and supermassive black hole properties of galaxies at high-angular resolution and directly probes the active galactic nuclei (AGN) hosted by the most massive quiescent galaxies deemed crucial for galaxy formation (D. A. Evans et al. 2006; V. Smolčić & D. A. Riechers 2011; V. Smolčić et al. 2017).

The three main contributors the extragalactic radio continuum emission at 1.4 GHz are (1) non-thermal synchrotron emission produced from AGN powered relativistic electrons interacting with large-scale galactic magnetic fields (H. Völk, U. Klein & R. Wielebinski 1989), (2) non-thermal synchrotron emission produced from cosmic ray electron produced from supernova remnants of massive ($M > 8\,M_{\odot}$) stars, and (3) thermal free–free emission from Coulomb scattering between free ions and electrons in H II regions ionized by young massive stars. The latter two are both tracers of star formation in galaxies (J. Condon 1992; E. Murphy et al. 2011). The source count population at 1.4 GHz is therefore a mixture of radio galaxies and quasars powered by AGN and star-forming galaxies (SFG). Radio Loud (RL) AGNs dominate the bright radio sky all the way down to the sub-mJy regime. At the sub-mJy regime, two populations are believed to be in competition, SFGs along with radio quiet (RQ) AGNs until below <100 μJy where SFGs become the most prominent radio sources making up the radio sky (A. Mignano et al. 2008; N. Seymour et al. 2008; V. Smolčić et al. 2008; M. Bonzini et al. 2013; P. Padovani et al. 2015; K. Kellermann et al. 2016; P. Padovani 2016).

★ E-mail: robert.kincaid@epfl.ch

The shape of the curves defined by the counts of radio sources as a function of their flux density was one of the earliest cosmological probes (S. Hoerner 1973; J. J. Condon 1988). Radio source counts continue to be of astrophysical interest as they are used to study the relative galaxy populations in the universe. The exact evolution of the Euclidean-normalized radio source counts and the relative radio galaxy populations and spectra are still a matter of debate. Current radio surveys are not sensitive enough to detect the fainter galaxies responsible for the bulk of star formation around the 'cosmic noon' at $z \sim 2$ (N. M. Förster Schreiber & S. Wuyts 2020). Most current samples are further hampered by their dependence on deep multiwavelength data covering small solid angles in fields selected using only optical/infrared criteria which are not ideal for making deep radio images (J. T. Zwart et al. 2014).

MeerKAT (J. L. Jonas 2009; J. Jonas & M. Team 2016) is a Square Kilometer Array (SKA) Precursor telescope that will eventually form part of the mid-frequency component of the SKA (D. B. Davidson 2012). The large number of baselines (2016), large collecting area ($64 \times 13.5$ m dishes), moderate field of view (FoV) (1 deg at *L*-band), and low (∼20 K) system temperature all make MeerKAT exceptionally fast at achieving image depth, which is important for galaxy continuum studies. Galaxy evolution continuum survey studies with MeerKAT that currently exist include The MeerKAT Galaxy Cluster Survey (K. Knowles et al. 2022) and MeerKAT International GHz tiered Extragalactic Exploration (MIGHTEE; M. J. Jarvis et al. 2017). These projects aim to explore galaxy star formation, galaxy evolution, AGN activity, and galaxy cluster dynamics.

Superclusters are the largest overdense non-virialized structures in the Universe. They extend over tens of megaparsecs and are composed of clusters, filaments and even small voids (J. Einasto, M. Jôeveer & E. Saar 1980; M. Einasto 2025). The evolution of galaxies residing inside superclusters are affected by the wide range of environments inside its host's supercluster environment (R. Seth & S. Raychaudhury 2020; I. G. Alfaro et al. 2022) thus making them excellent targets for study. Targeted study of galaxy supercluster regions with MeerKAT are scarce with only MeerKAT HI surveys existing for superclusters such as Virgo (G. Nagaraj, M. Ramatsoku & P. Jablonka 2024; A. Spasic et al. 2024) and Vela (R. C. Kraan-Korteweg et al. 2017). The MeerKAT Observations of the Saraswati Supercluster (MOSS) project therefore gives us a unique opportunity to test the capabilities of MeerKAT in a new realm of deep continuum survey studies of superclusters.

### 1.1 Saraswati supercluster

The *Saraswati* supercluster was discovered by J. Bagchi et al. (2017) in the Stripe 82 region of Sloan Digital Sky Survey (SDSS-III DR12) SDSS (S. Alam et al. 2015). The existence of *Saraswati* is further supported by S. Sankhyayan et al. (2023), who found its mass to be comparable to that of low-redshift massive superclusters such as the Sloan Great Wall (J. R. Gott III et al. 2005) and Corona Borealis (M. Einasto et al. 2021).

Located at $z = 0.28$, extending $\sim 200$ Mpc across in comoving coordinates and with a mass of about $2 \times 10^{16}\,\mathbf{M}_{\odot}$, *Saraswati* stands out in the sky as an especially rare and possibly one of the largest significant density enhancements found at intermediate redshifts. There are about 43 massive galaxy clusters or groups in this Supercluster region with five high-mass galaxy clusters concentrated in its dense (gravitationally bound) core region. This core region extends to a radius of 20 Mpc and encompasses a mass of at least $4 \times 10^{15}$ $M_{\odot}$ (20 per cent of total supercluster mass). These properties would suggest that it was probably formed by some extreme, large-scale, matter density enhancements in the early universe around ∼4 Gyr. The most and second most massive galaxy clusters are the clusters Abell 2631 (R. Monteiro-Oliveira et al. 2021) (A2631 hereafter) and ZwCL2341.1 + 0000 (J. Bagchi et al. 2002) (ZwCL2341 hereafter), situated near the core of *Saraswati*. They are surrounded by a filamentary network of galaxies. These two clusters form the focal point of this paper.

The outline of the paper is as follows: In Section 2, we describe briefly the previous data-reduction and calibration steps that produced the final images used for this paper. We then present these final images before and after primary beam correction cutoff in Section 3 and describe the source extraction procedure in generating our final catalogues. The noise distribution and visibility area of the cutoff final images are displayed in Section 4, we also include checks of total flux scale, astrometry accuracy and source classification for each catalogue. In Section 5, we derive the radio spectral index distribution for each catalogue and look into trends of spectral index with total (integrated) flux density. Biases that result in the incompleteness of our radio catalogues are explored, namely the noise bias (through simulations), resolution bias, and Eddington bias (in Section 6). This is then followed by Section 7 where we derive the intrinsic 1.4 GHz source counts for the two catalogues and compare with existing evolutionary models and MeerKAT data. Lastly, we summarize and conclude our results in Section 8.

## 2 RADIO OBSERVATION AND DATA REDUCTION

The details regarding observation, data reduction, calibration, and imaging procedure for the clusters A2631 and ZwCL2341 are described in detail in the initial paper MOSS 1 of V. Parekh et al. (2022). We show a summary of the observation details in Table 1 and highlight a few of the data-reduction and calibration steps here.

Each cluster was observed with the MeerKAT telescope for 7 h with 4k channels and 856 MHz bandwidth using 60 out of the total 64 MeerKAT antennas, with the MeerKAT standard primary and secondary calibrators (See Table 1). Data reduction was performed with the CARACAL pipeline[1]. V. Parekh et al. (2022) used the CARACAL pipeline for direction independent self calibration and imaging with CUBICAL[2] (J. S. Kenyon, S. Perkins & O. Smirnov 2023) and WSCLEAN (A. Offringa & O. Smirnov 2017), respectively. The WSCLEAN Multi-Frequency Synthesis (MFS) images produced after self-calibration contained significant residual calibration errors, that manifest themselves as error patterns resembling a corrupted point spread function (PSF) around off-axis bright (∼ tens of mJy or stronger) sources. These type of sources are termed direction-dependent effects (DDEs).

[1] https://github.com/caracal-pipeline/caracal
[2] The CUBICAL software is currently called QUARTICAL, found at https://github.com/ratt-ru/QuartiCal

**Table 1.** MeerKAT observations of the core region of *Saraswati* supercluster.

| Instrument: | MeerKAT |
|---|---|
| Observation date: | 2019-06-15 |
| Number of pointings: | 2 |
| Number of antennas: | 60 |
| Total observation time: | 14 h |
| Central frequency: | 1283 MHz |
| Total bandwidth: | 900 MHz |
| Channel width: | 208 kHz |
| Total number of channels: | 4016 |
| Integration time: | 16 s |
| Cross products: | XX, XY, YX, YY |
| Band pass and flux calibrator: | J1939−6342 |
| J1939−6342 flux density at 1.4 GHz: | ∼14.90 Jy |
| Gain calibrator: | J2357−1125 |
| J2357−1125 flux density at 1.4 GHz: | ∼1.80 Jy |
| A2631 RA, Dec: | 354.4191, 0.2766 |
| ZwCL2341 RA, Dec: | 355.9154, 0.3308 |

### 2.1 Direction-dependent calibration

At *L*-band, DDEs are caused by the baseline, time, frequency, and direction dependent variations in the complex antenna beam pattern, coupled with pointing errors (K. Asad et al. 2021). V. Parekh et al. (2021) explored different DD calibration methods used for reduction of bright problematic sources residing in the FoV of A2631 and ZwCl2341. Here we outline the details of one such method used in their paper. We chose the facet-based approach of the KILLMS[3] solver (C. Tasse 2014b; O. Smirnov & C. Tasse 2015) in combination with DDFACET imager (C. Tasse et al. 2018) to minimize the contamination from DDEs.

In total, four runs of DDFACET imaging and a single run of KILLMS for the DD calibration was performed. During the DDFACET imaging stage, we progressively increase the number of iterations in each cleaning cycle and use deeper masks to improve the sky model. The model created after self calibration was partitioned into ∼7 directions depending on the location of bright sources. This partitioning forms the tessellated structure shown in Fig. 1[4] and is used by DDFACET for the facet-based cleaning approach. The DDFACET imaging runs used J. Högbom (1974) CLEAN for the deconvolution of each facet in five frequency sub-bands. A briggs robustness parameter value of 0 was used for an image optimized for sensitivity. The initial DDFACET run created the first order skymodel and its resulting image was used to create a shallow mask of threshold 20. Subsequent DDFACET runs cleaned within this mask and the resulting imaging products were used to create deeper masks until a threshold of 5 was reached. The final DDFACET run cleaned within the deep mask resulting in a deep clean skymodel.

[3] https://github.com/cyriltasse/killMS
[4] This tesselation was created using the `makeModel.py` script of https://github.com/saopicc/DDFacet/tree/master/SkyModel

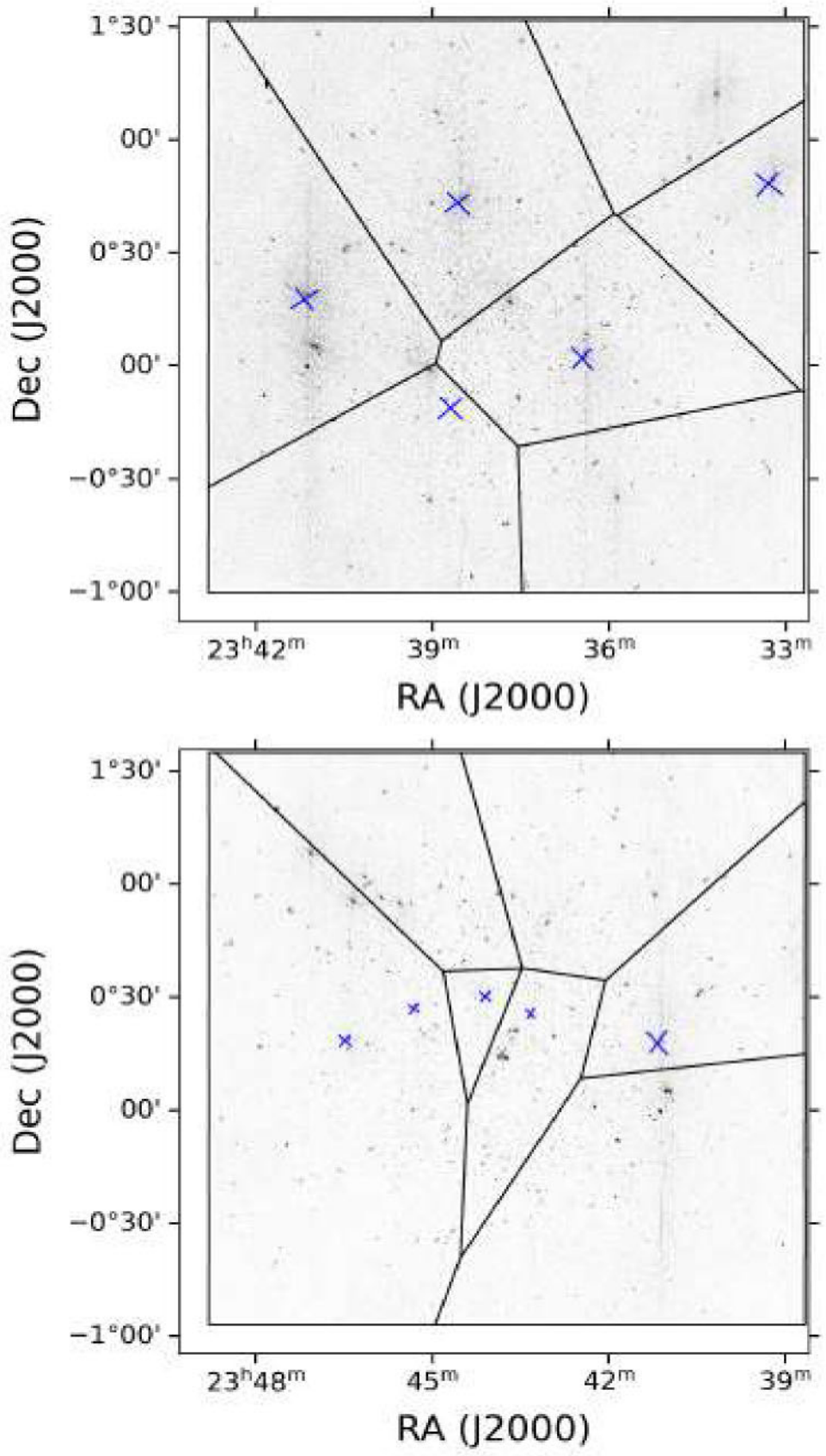

**Figure 1.** Tesselation of the image created from the location of off-axis bright sources for A2631 (top) and ZwCL2341 (bottom). These off-axis bright sources are marked by the crosses and exhibit long streak-like vertical noise patterns. KILLMS solves for complex gain solutions for each tessel independently. These solutions are then applied in DDFACET during the deconvolution of each facet.

This deep skymodel was then used in KILLMS to solve for complex gain corrections in directions defined by the tesselation of the image (see Fig. 1). We used the non-linear Kalman Filter algorithm (KAFCA; C. Tasse (2014a)) to solve for each direction defined by the tesselation with a time/frequency interval of 5 min/20 channels. A final fourth DDFACET run applied these solutions through the option `-DDESolutions-DDSols` using the same deep mask and with 200 000 minor iterations for the deconvolution.

### 2.2 Primary beam correction and cutoff

V. Parekh et al. (2021) used the facet-based primary beam correction during the DDFACET imaging that are applied on a per-tesselation basis in much the same way the DD gain corrections are applied. The beam model consists of a multidimensional FITS model that contains a full $2 \times 2$ Jones matrix model of the MeerKAT primary beam as a function of direction, evaluated at several frequency intervals. The primary beam correction is applied on each visibility in the form of convolution kernel during gridding (S. Bhatnagar, U. Rau & K. Golap 2013). The advantage of this visibility-based approach of primary beam correction is that the frequency, time, polarization, and direction dependences of the primary beam can be fully modelled.

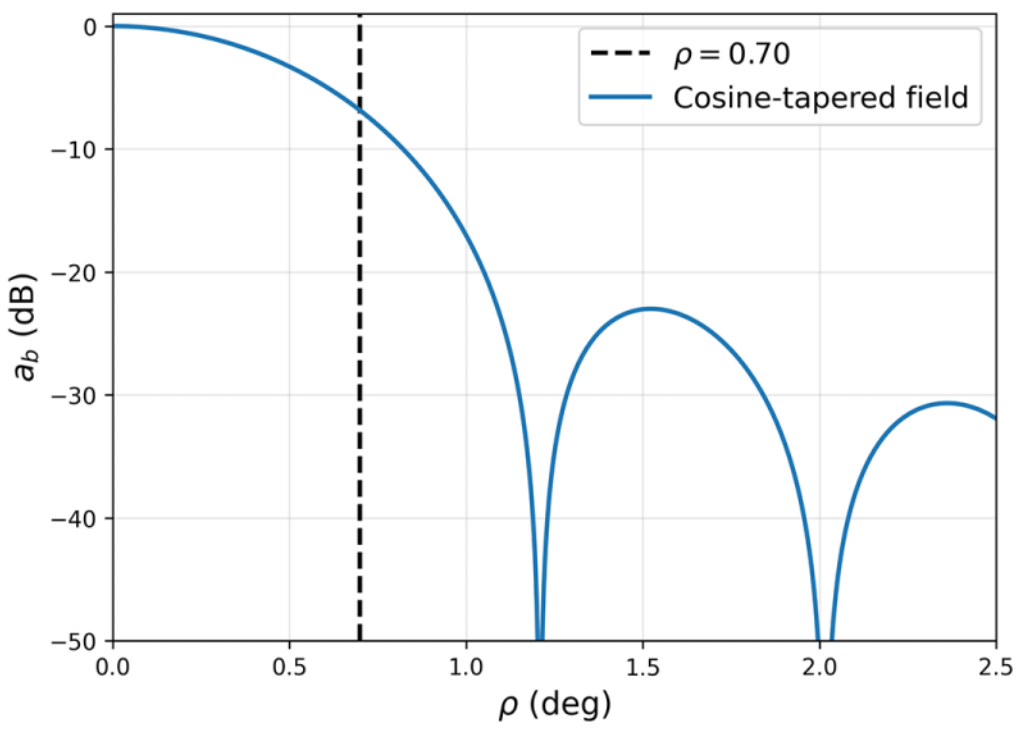

**Figure 2.** The simplified model of KATBEAM modelled using the cosine-tapered field at full width half maximum (FWHM) $\theta_b = 57.5$ arcmin shown by the solid curve. The vertical line shows $\rho = 0.7$ deg, the radius of the resulting image after masking the image where the primary beam attenuation of KATBEAM model (black curve) exceeds 30 per cent. From Figure 4 of T. Mauch et al. 2020, the cosine-tapered model of KATBEAM at $\rho < 0.7$ essentially matches real measurements of the 1.5 GHz Stokes I primary beam power pattern.

After primary beam correction, the noise of an image increases as a function of distance from the phase centre. We chose to cutoff our image in regions where the noise exceeds a certain value. Due to the complexity of the resulting primary beam pattern we made use of the simplified MeerKAT beam model of KATBEAM to determine the regions in which to perform the primary beam cutoff. The KATBEAM model attenuation pattern uses the cosine-tapered field illumination (J. J. Condon & S. M. Ransom 2016) and is a good approximation to real measuremnts of the MeerKAT primary beam (See Figure 4 of T. Mauch et al. 2020). We show in Fig. 2 a plot of the cosine-tapered field used by KATBEAM model (solid curve). We chose to mask our images out to 0.3 times the beam response (≈30 per cent attenuation), which is less than the half-power radius. This resultant masked image had a new area of ∼1.6 deg$^2$ and a corresponding radius of ∼0.7 deg. The vertical dashed line in Fig. 2 corresponds to $\rho = 0.7$ deg which is well within the MeerKAT main lobe.

## 3 FINAL IMAGE AND CATALOGUE CONSTRUCTION

### 3.1 Final image

The full primary beam corrected and KATBEAM cutout DD-calibrated DDFACET MFS images of A2631 (top) and ZwCL2341 (bottom) are shown on the left and right sides in Fig. 3, respectively. On top of this figure we show in Table 2 properties pertaining to each cluster such as number of sources, Root Mean Square (RMS) noise, area of full and cutout images, etc. The cutout images are the masked versions of the full images where the primary beam response exceeded that of 30 per cent in an attempt to mask the noisy side lobes and only consider sources located within the primary beam main lobe. We show in Fig. 4 examples of extended and compact sources existing within the cutout images.

### 3.2 Catalogue construction

We used the Python Blob Detection and Source Finder PYBDSF[5] (N. Mohan & D. Rafferty 2015) to produce radio catalogues of the two clusters. PYBDSF searches for islands of emission at a threshold above the image mean and then attempts to fit components composed of one or more overlapping elliptical gaussians within those islands with a peak pixel above a certain threshold. The fitted gaussians are then used to determine the flux densities, source sizes, positions, and source type. The number of gaussian components grouped and fitted to a single island determines the source type. Three types of classifications can result from the fitting of gaussians to these islands of emissions: (1) A single Gaussian for an island classified as 'S', (2) a single Gaussian that overlaps with an island containing other gaussians classified as 'C', and (3) multiple Gaussian fits for a single island classified as 'M'. These source types are assigned under the column `Source_Code` in the catalogue.

The images inserted in PYBDSF for radio source catalogue extraction were the MeerKAT primary beam corrected cutout images generated from the DDFACET software shown on the right panels of Fig. 3 . For the PYBDSF source extraction procedure we invoked the task `process_image`, with the parameters of a peak detection threshold of `thresh_pix` = 5 and an island detection threshold of `thresh_isl` = 4.

The images present several strong sources exhibiting strong artefacts resulting from incomplete deconvolution of the PSF due to calibration errors. These artifacts can be falsely classified as sources with PyBDSF if we do not constrain the island threshold around these problematic regions. This can be done using a dynamic RMS estimation method that uses an RMS box of different size depending on flux distribution to accurately capture the RMS variation. This is done by first selecting a fixed RMS box size `rms_box` and then decreasing the box size in regions over which the brightness of the artefacts increase significantly using `bright_box`. By having a smaller box size around the bright artefacts, we are able to fit islands around real emission and not residual noise. We tried a variety of sliding box sizes and found that a slight adjustment to the method used in W. Williams et al. (2021) worked well. We found that an RMS box of 40 × 40 pixels (`rms_box = (40,40)`) to estimate the RMS noise across the image and then decreased it to 20 × 20 (`rms_box = (20,20)`) for regions of signal noise ratio (S/N) > 150 (`adaptive_threshold = 150`) worked the best.

The excerpts of these catalogues are shown in Tables 3 and 4 for A2631 and ZwCL2341, respectively. A total of 2611 sources were found for ZwCL2341 with total fluxes densities between 23 μJy and 0.5 Jy. For A2631. A total of 1999 sources were found with total flux densities between 30 μJy and 0.25 Jy. In Fig. 5, the RA and DEC distributions of all radio sources in these catalogues are shown for A2631 (blue) and ZwCL2341 (red). The darker colours indicate sources found in the the cutout images and lighter colours represent the sources found when using the full images (left i, 132, 03500mages of Fig. 3) using the same PYBDSF parameters used in cutout images.

## 4 IMAGE ANALYSIS

In this section, we investigate the noise properties from the output RMS images produced from source extraction (Section 3.2). We then check the consistencies of the flux scale in Section 4.2 by comparing the total flux densities to the FIRST radio survey. We also check the astrometry of our sources with the VLASS radio survey in Section 4.3 and then the fraction of resolved and unresolved sources are performed in Section 4.4.

[5]https://github.com/lofar-astron/PyBDSF

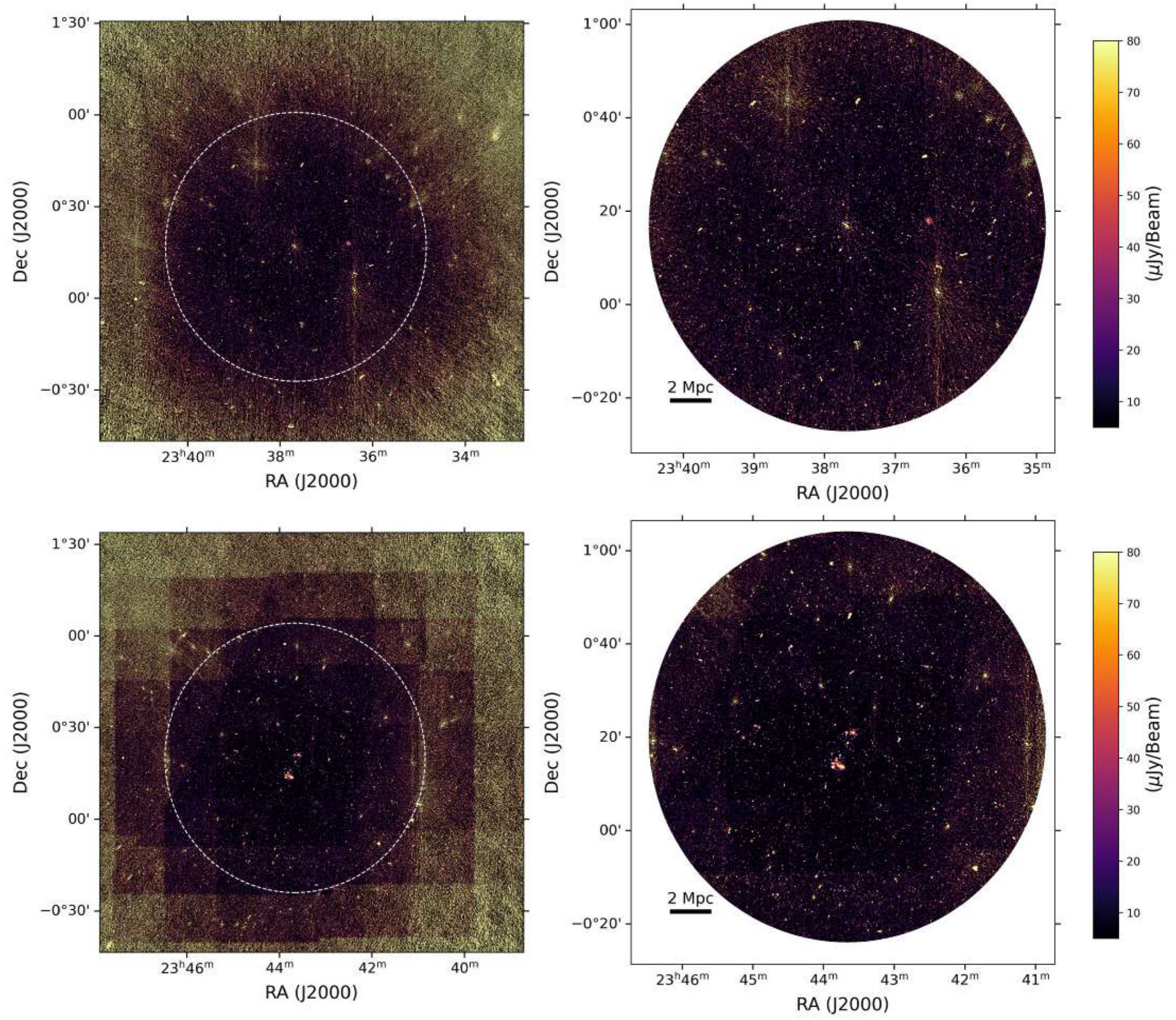

**Figure 3.** Left: full-field ∼6 deg$^2$ primary beam-corrected image of A2631 (top) and ZwCL2341 (bottom). Right: ∼1.6 deg$^2$ cutout image resulting from masking the image on the left at a distance from pointing centre of 0.7 deg of the MeerKAT primary beam power (cutting off regions with > 30 per cent attenuation) for A2631 (top) and ZwCL2341 (bottom). This cutout region is outlined by the dashed white ellipse on the left images. We show at the bottom left the angular scale equivalence of 2 Mpc at the cluster redshift. Colour scale intensity for all plots is between $0.5\sigma$ to $8\sigma$, where $\sigma = 10, 15$ µJy beam$^{-1}$ for A2631 and ZwCL2341, respectively.

### 4.1 Noise analysis

Radio images usually have a varying sensitivity over the region surveyed. This means that the area over which a source can be detected (visibility area) increases with peak flux density. We can evaluate the spatial variation of the sensitivity using the noise map created by PyBDSF obtained from specifying `img_type = 'rms'` option when running the `export_image` task. This noise map is shown in Fig. 6. The noise varies from $\sigma \sim 10$ µJy beam$^{-1}$ in the central region to about ∼60 µJy beam$^{-1}$ at the edges where the primary beam cutoff was imposed. At the location of bright sources the noise can increase to values $\sim 10 \times \sigma$. This is due to the presence of residual amplitude or phase errors present after calibration.

The total area in which a source of a given flux density can be detected (visibility area) is shown in Fig. 7. For both clusters, the visibility area shows a rapid increase from the central RMS value of $\sigma \sim 10$ µJy beam$^{-1}$ (∼1 per cent of image area) to ∼40 µJy beam$^{-1}$ (∼99 per cent of image area) and then gradually increases until 100 µJy beam$^{-1}$. Approximately 99 per cent of the area in both images have an RMS noise less than 60 µJy beam$^{-1}$.

**Table 2.** Statistical properties of the core galaxy clusters in the *Saraswati* supercluster.

| Field | RA (hh:mm:ss) | DEC (hh:mm:ss) | $\sigma$(central RMS) μJy beam$^{-1}$ | Area$^{\mathrm{full}}$ deg$^2$ | $N^{\mathrm{full}}$ ($N$ sources) in full map) | Area$^{\mathrm{masked}}$ (deg$^2$) | $N^{\mathrm{cutout}}$ ($N$ sources in cutout map) | flux density min (mJy) | flux density max (mJy) |
|---|---|---|---|---|---|---|---|---|---|
| A2631 | 23 : 37 : 40.60 | +00 : 16 : 36.0 | 10 | 6.1 | 3328 | 1.6 | 1999 | 0.038 | 250 |
| ZwCL2341 | 23 : 43 : 39.70 | +00 : 19 : 51.0 | 15 | 6.1 | 3437 | 1.6 | 2481 | 0.023 | 500 |

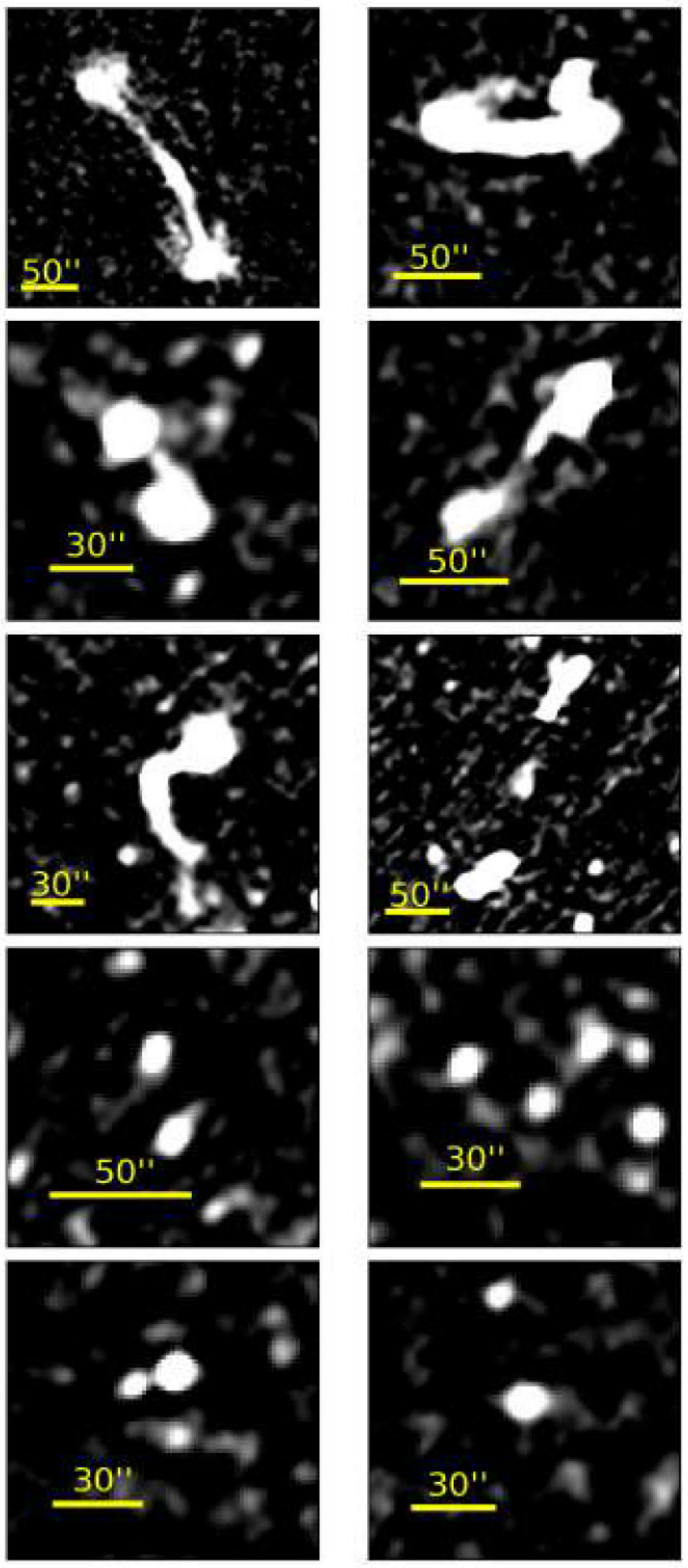

**Figure 4.** Cutouts from A2631 and ZwCL2341 images showing examples of detected extended sources (first six cutouts) and compact sources (last four cutouts) present in both fields.

## 4.2 Flux density accuracy

Observations from different instruments have their own intrinsic primary beam models and flux calibrators. We therefore expect some systematic offsets in the measured fluxes of MeerKAT with other radio instruments. We have chosen to use FIRST (R. L. White et al. 1997) survey catalogues to compare the MeerKAT total flux densities. FIRST with its similar frequency of observation (1.4 GHz), angular resolution (∼5 arcsec) and good coverage of the regions hosting clusters A2631 and ZwCL2341 make it a suitable survey for MeerKAT flux density measurement comparison.

FIRST image cutouts were obtained from SKYVIEWER[6] for the central area out to a radius of 1 deg field of both A2631 and ZwCL2341. We then convolved each FIRST image with the larger MeerKAT beam of ∼10 arcsec and regridded the MeerKAT images to the finer pixel grid of the FIRST image cutouts to ensure both images are set on the same pixel scale. After these two operations, we ran PYBDSF on the MeerKAT regridded images and FIRST convovled images of A2631 and ZwCL2341. We then cross-matched the resulting catalogues using a separation

[6] https://skyview.gsfc.nasa.gov/current/cgi/query.pl

**Table 3.** A sample of ten rows from the source catalogue of A2631.

| Source ID | Name | RA (J2000, deg) | E_RA (J2000, deg) | DEC (J2000, deg) | E_DEC (J2000, deg) | Maj (arcsec) | Min (arcsec) | Total_flux (μJy) | E_Total_flux (μJy) | Peak_flux (μJy) | E_Peak_flux (μJy) | RMS (μJy beam$^{-1}$) | – | S_code |
|---|---|---|---|---|---|---|---|---|---|---|---|---|---|---|
| (1) | (2) | (3) | (4) | (5) | (6) | (7) | (8) | (9) | (10) | (11) | (12) | (13) | | (15) |
| 0 | A2631 J234026.97 + 001803.0 | 355.112367 | 0.00023 | 0.300837 | 0.00026 | 11.19 | 7.22 | 369 | 118.47 | 287 | 56.38 | 55.50 | – | S |
| 1 | A2631 J234023.32 + 002631.0 | 355.097154 | 0.00014 | 0.441931 | 0.00011 | 8.55 | 7.02 | 310 | 73.20 | 325 | 43.26 | 44.09 | – | C |
| 2 | A2631 J234023.94 + 002629.5 | 355.099738 | 0.00015 | 0.441529 | 0.00015 | 7.57 | 7.40 | 228 | 68.66 | 256 | 42.74 | 44.09 | – | C |
| 3 | A2631 J234019.92 + 000602.0 | 355.082999 | 0.00026 | 0.100557 | 0.00018 | 17.47 | 13.52 | 1477 | 215.62 | 393 | 46.30 | 45.58 | – | S |
| 4 | A2631 J234019.98 + 002634.1 | 355.083233 | 0.00029 | 0.442812 | 0.00028 | 14.19 | 8.65 | 520 | 139.69 | 266 | 49.53 | 47.40 | – | S |
| 5 | A2631 J234018.77 + 000612.8 | 355.078203 | 0.00012 | 0.103566 | 0.00015 | 10.04 | 8.20 | 547 | 106.65 | 418 | 50.99 | 48.68 | – | S |
| 6 | A2631 J234017.95 + 001406.7 | 355.074807 | 0.00001 | 0.235190 | 0.00001 | 8.68 | 8.24 | 5015 | 90.81 | 4408 | 48.05 | 46.69 | – | S |
| 7 | A2631 J234018.39 + 000831.3 | 355.076638 | 0.00003 | 0.142028 | 0.00003 | 8.59 | 7.56 | 1586 | 85.41 | 1536 | 48.13 | 47.90 | – | S |
| 8 | A2631 J234017.25 + 002135.9 | 355.071890 | 0.00014 | 0.359978 | 0.00013 | 10.15 | 7.63 | 406 | 82.36 | 329 | 40.84 | 39.67 | – | S |
| 9 | A2631 J234015.21 + 003216.8 | 355.063393 | 0.00001 | 0.537993 | 0.00001 | 0.00 | 0.00 | 27 188 | 302.39 | 27 188 | 173.29 | 173.12 | – | S |
| 10 | A2631 J234014.07 + 003014.9 | 355.058620 | 0.00011 | 0.504127 | 0.00011 | 10.00 | 9.06 | 913 | 138.91 | 633 | 62.30 | 58.39 | – | S |

**Table 4.** A sample of ten rows from the source catalogue of ZwCl2341.

| Source ID | Name | RA (J2000, deg) | E_RA (J2000, deg) | DEC (J2000, deg) | E_DEC (J2000, deg) | Maj (arcsec) | Min (arcsec) | Total_flux (μJy) | E_Total_flux (μJy) | Peak_flux (μJy) | E_Peak_flux (μJy) | RMS (μJy $beam^{-1}$) | – | S_code |
|---|---|---|---|---|---|---|---|---|---|---|---|---|---|---|
| (1) | (2) | (3) | (4) | (5) | (6) | (7) | (8) | (9) | (10) | (11) | (12) | (13) | | (14) |
| 0 | ZwCL2341 J234625.83 + 001604.2 | 356.607633 | 0.00004 | 0.267830 | 0.00003 | 15.47 | 9.61 | 7321 | 268.49 | 2744 | 54.25 | 54.25 | – | M |
| 1 | ZwCL2341 J234626.46 + 001611.7 | 356.610248 | 0.00004 | 0.269919 | 0.00004 | 10.87 | 8.78 | 2603 | 134.87 | 1703 | 57.85 | 54.25 | – | C |
| 2 | ZwCL2341 J234624.61 + 001914.4 | 356.602525 | 0.00000 | 0.320677 | 0.00000 | 0.00 | 0.00 | 67 166 | 634.70 | 67 166 | 299.45 | 299.45 | – | M |
| 3 | ZwCL2341 J234624.89 + 001836.2 | 356.603693 | 0.00005 | 0.310048 | 0.00006 | 9.68 | 8.77 | 4942 | 360.44 | 3635 | 168.48 | 159.09 | – | S |
| 4 | ZwCL2341 J234624.64 + 002033.1 | 356.602673 | 0.00011 | 0.342535 | 0.00012 | 7.08 | 5.44 | 173 | 51.80 | 281 | 39.99 | 47.71 | – | S |
| 5 | ZwCL2341 J234624.65 + 002055.7 | 356.602699 | 0.00018 | 0.348816 | 0.00015 | 8.09 | 7.01 | 200 | 66.84 | 221 | 40.99 | 42.23 | – | S |
| 6 | ZwCL2341 J234624.54 + 001756.3 | 356.602232 | 0.00009 | 0.298982 | 0.00011 | 6.44 | 5.25 | 110 | 34.63 | 203 | 29.11 | 36.63 | – | S |
| 7 | ZwCL2341 J234623.38 + 002052.3 | 356.597431 | 0.00022 | 0.347863 | 0.00017 | 0.00 | 0.00 | 201 | 68.23 | 201 | 39.26 | 40.06 | – | S |
| 8 | ZwCL2341 J234623.24 + 003029.6 | 356.596816 | 0.00023 | 0.508211 | 0.00024 | 9.63 | 9.07 | 217 | 71.02 | 155 | 32.55 | 30.59 | – | S |
| 9 | ZwCL2341 J234623.03 + 001501.9 | 356.595951 | 0.00011 | 0.250525 | 0.00012 | 9.00 | 7.98 | 530 | 99.60 | 461 | 52.24 | 50.77 | – | S |
| 10 | ZwCL2341 J234622.77 + 002518.1 | 356.594895 | 0.00014 | 0.421704 | 0.00013 | 0.00 | 0.00 | 297 | 75.53 | 297 | 42.72 | 42.44 | – | S |

*Note.* (1) Source unique identifier, (2) source name, (3, 4) source right ascension and error, (5, 6) Source declination and error, (7, 8) deconvolved major and minor axis size, number of sources, (9, 10) total flux density and error, (10, 11) peak flux density and error, (13) local RMS, and (14) source classification code based on the number of Gaussians fitted per island. The three possible classifications are ‘S’, ‘C’, ‘M’.

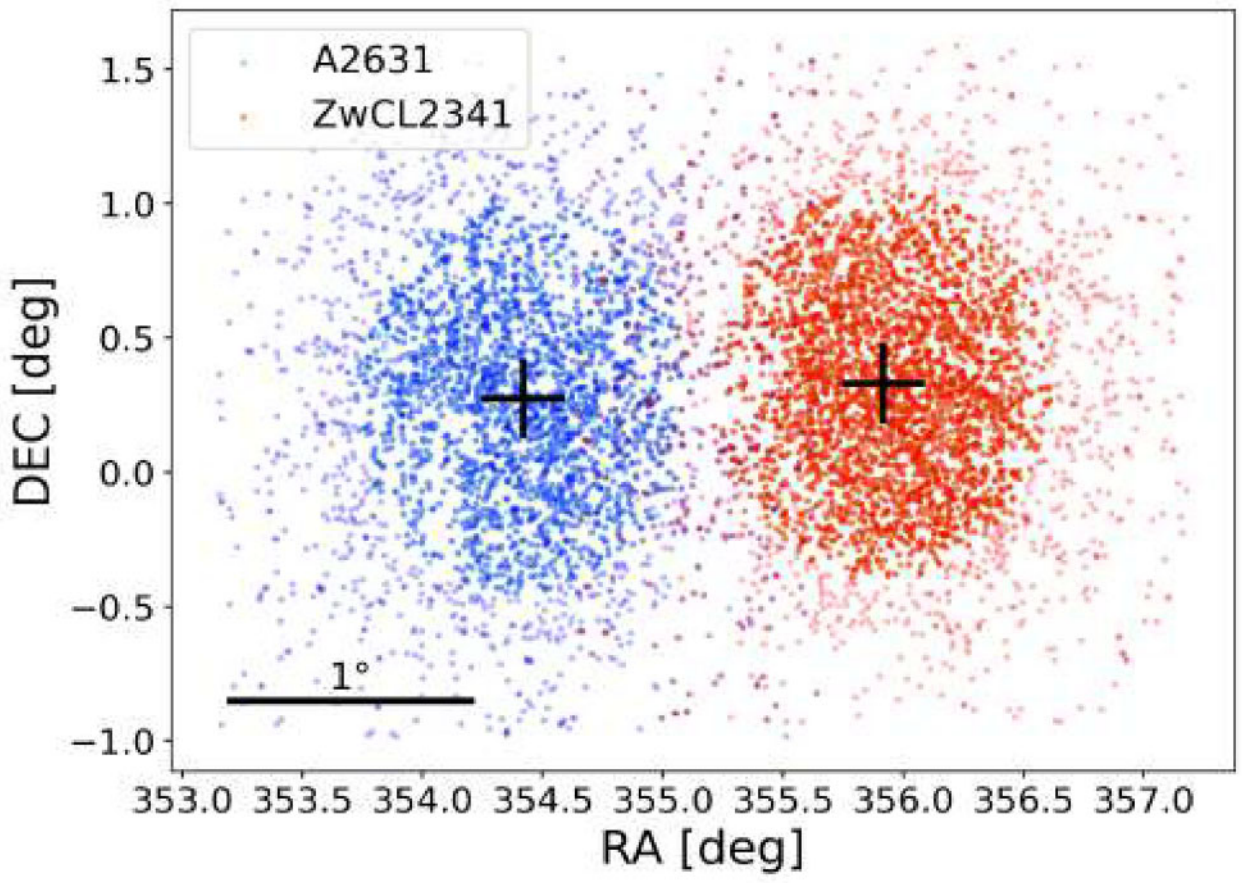

**Figure 5.** RA Dec. distributions for sources in A2631 (blue) and ZwCL2341 (red) that reside in the *Saraswati* core region. The green crosses in the middle represent the pointing centres. The lighter colours indicate sources found in the full image (left images of Fig. 3) and the darker colours represent sources found in the images clipped where the primary beam attenuation exceeds that of 30 per cent (right images of Fig. 3). These darker colour sources represent the sources in our final catalogue used for this paper.

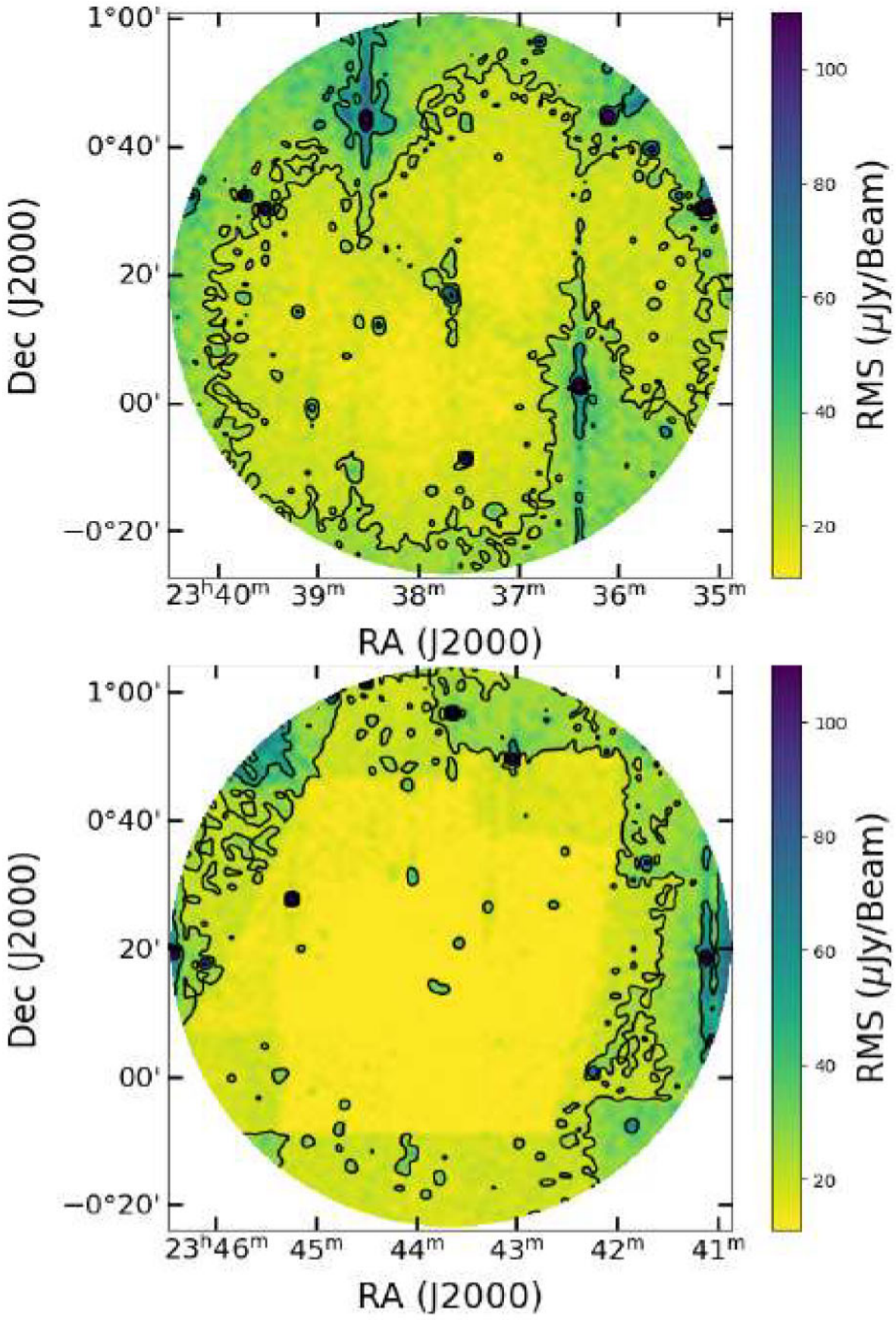

**Figure 6.** Top: RMS noise map of A2631 (top) and ZwCL2341 (bottom) from PyBDSF. The colour scale varies from $1\sigma$ to $10\sigma$, where $\sigma = 16$, 10 μJy for A2631 and ZwCL2341, respectively. Contours are plotted at $2^i \times \sigma$ where $i = [1, 2, 3, 4, 5, 6]$. Peaks in the RMS noise correspond to locations of the bright sources or calibration artefacts.

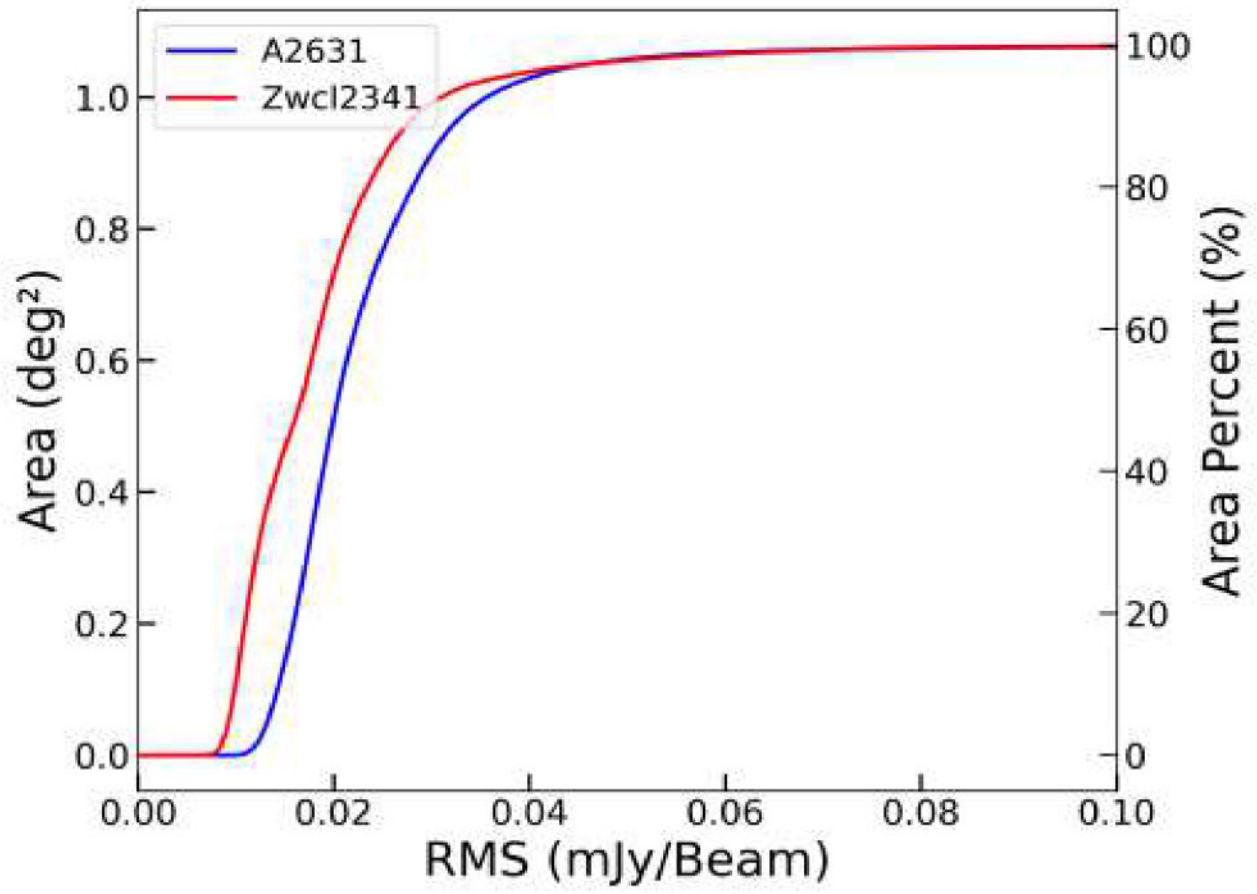

**Figure 7.** Visibility area of A2631 and ZwCL2341. The visibility function represents the cumulative fraction of the total area of the noise map characterized by a noise lower than a given value.

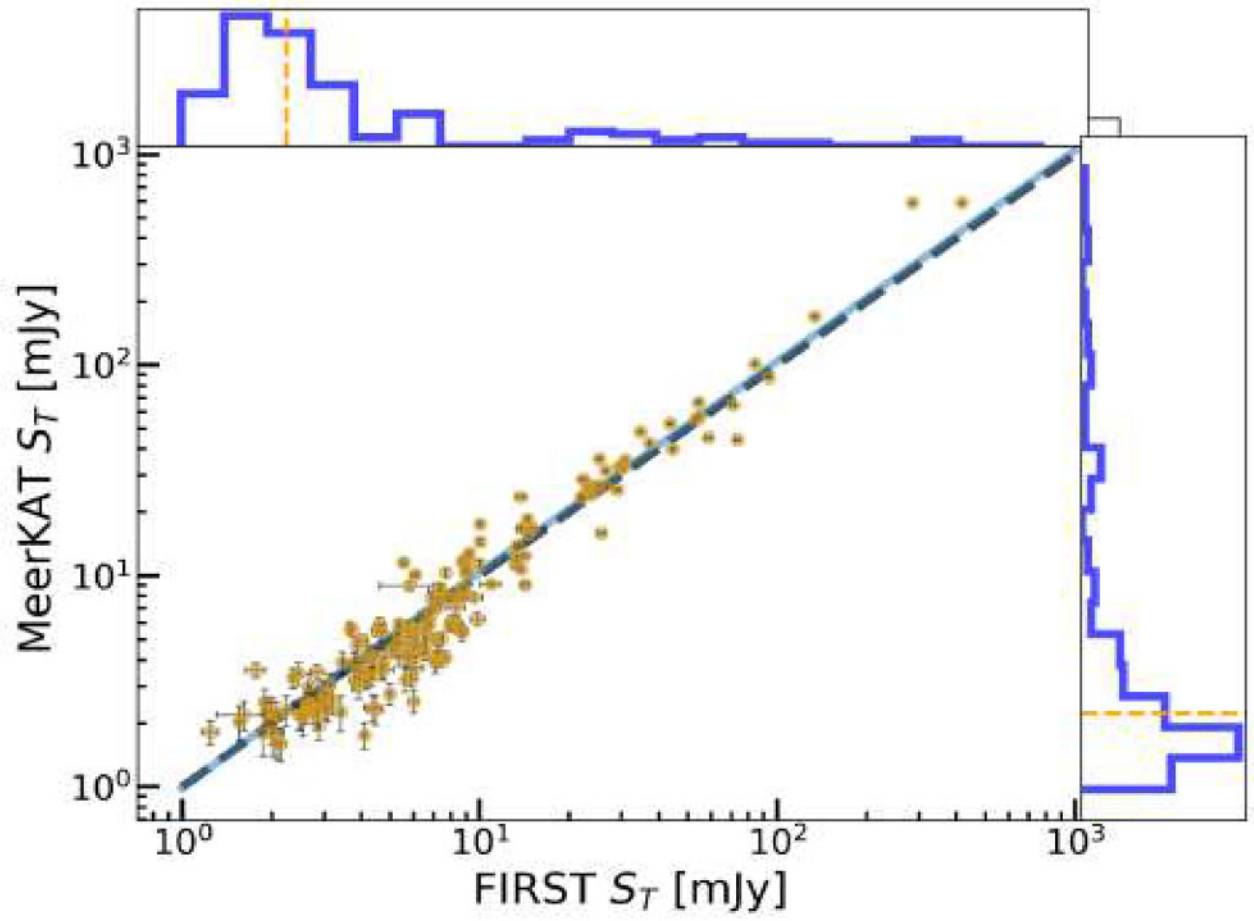

**Figure 8.** Total flux density comparison at 1.4 GHz for the MeerKAT/FIRST combined sources of A2631 and ZwCl2341. We have only selected high S/N (S/N $>$ 10) and unresolved and sources. The dashed line indicated a flux density ratio equal to unity. The transparent line indicates the best fitted line (with $R^2 = 0.98$) to the data. Histograms of the flux density distribution are shown horizontally on the top and sideways on the vertical axis for MeerKAT and FIRST, respectively. The dashed line in these panels represents the median flux density.

(tolerance radius) of 5 arcsec, yielding a combined total of ∼800 MeerKAT/FIRST sources for A2631 and ZwCL2341.

The following criteria were then applied to these sources in an attempt to obtain a suitable sample of sources for total flux density comparison: (1) high S/N sources with S/N $>$ 10 and (2) unresolved compact sources with peak flux density $S_{\rm P}$ and total flux density $S_{\rm T}$ between $0.8 < S_{\rm T}/S_{\rm P} < 1.2$. This reduced the size of the cross-matched catalogue to 675 sources. Lastly, we scaled the MeerKAT fluxes to 1.4 GHz using a spectral index of $\sim -0.7$.

We have plotted the total flux density comparison of FIRST with MeerKAT in Fig. 8 between the flux density range $1 < S_{\rm T} <$ 1000mJy. We show the 1:1 line in dashed black and the best-fitting curve in blue. The errors in the MeerKAT and FIRST total flux density measurement are reported by PyBDSF in the `E_Total_flux` column. We see that from Fig. 8, the flux den-sities of the FIRST versus MeerKAT sources are generally well-correlated. This is especially true for the brighter sources where the flux density ratio correlation is tighter with the 1:1 dashed black line. For the lower end containing fainter sources, there is more scatter. We used simple linear regression to fit a curve between the FIRST and MeerKAT total flux densities. This best-fitting curve shown in blue and has a correlation strength of $R^2 = 0.98$. We find a mean flux density ratio (MeerKAT $S_T$/FIRST $S_T$) of $f_r = 1.11$ and standard deviation of $\sigma_{f_r} = 0.36$ for our sample of sources. This flux density ratio indicates systematic offset of 11 per cent in our MeerKAT flux scale in comparison with FIRST. This offset was not applied to the total flux density density of our sources in the catalogue.

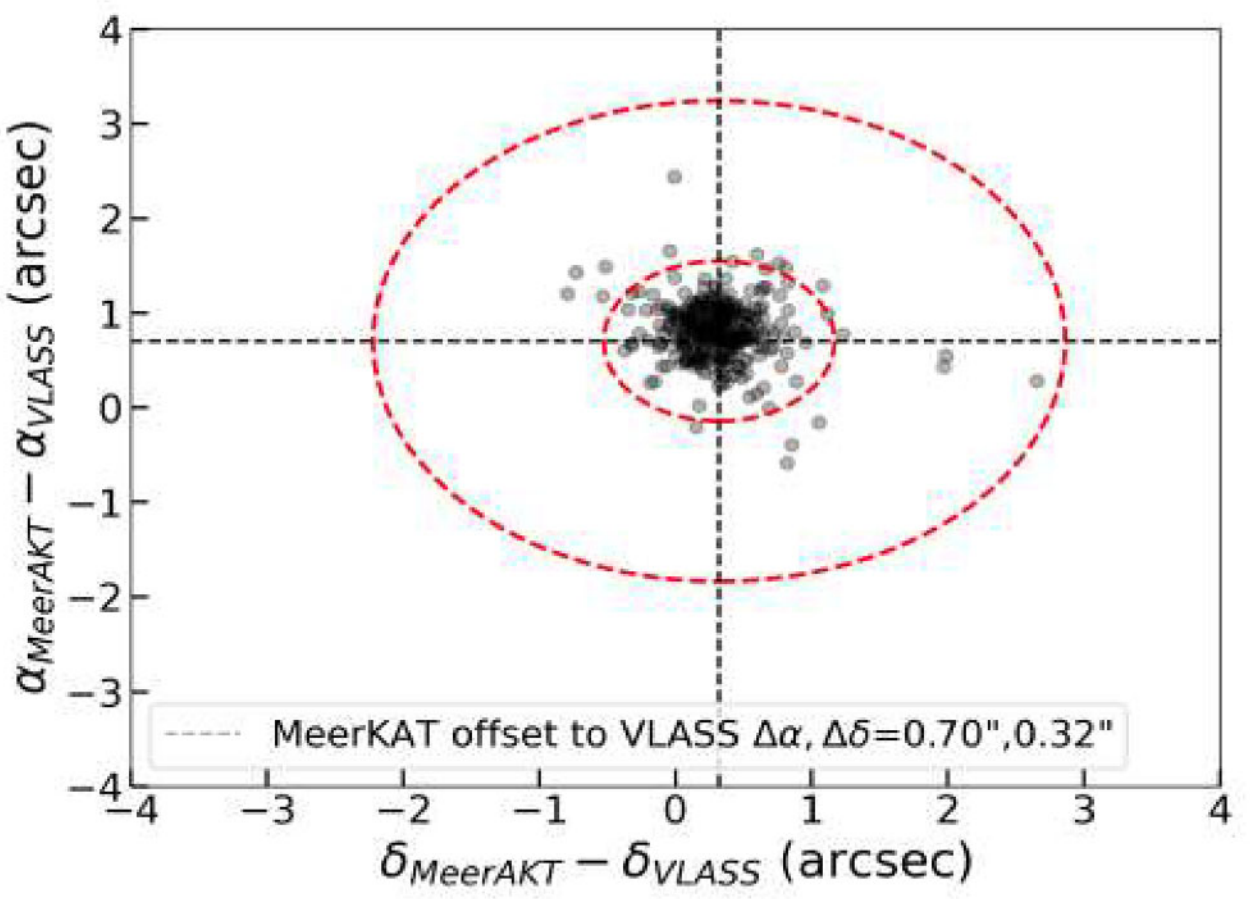

**Figure 9.** The combined offset in RA and Dec. between MeerKAT with VLASS for the sources of A2631 and ZwCL2341. The inner red circle and outer red circle corresponds to $1\sigma$ and $3\sigma$, respectively, where $\sigma$ is the standard deviation in offset. The horizontal and vertical dotted lines correspond to the mean in the (RA $\Delta\alpha = 0.70$ arcsec) and Dec. ($\Delta\delta = -0.32$ arcsec), respectively.

We have additionally made use of a high-resolution deep 1.6 GHz image of A2631 from the Expanded VLA (EVLA) as a consistency check for the MeerKAT fluxes. This analysis is shown in Appendix B.

### 4.3 Astrometry

We evaluated the source positional offsets induced by phase calibration errors by comparing our MeerKAT positions to that of VLASS (resolution ~2.5 arcsec, RMS = 70 μJy). See Y. A. Gordon et al. (2021) for specifications and source statistics of VLASS. We chose to compare with the VLASS due to their good positional accuracy (<0.5 arcsec) in the southern sky ( > −20 deg) and the large number of VLASS sources situated in the vicinity of A2631 and ZwCl2341. We found a total of 1020 and 981 VLASS sources for A2631 and ZwCL2341, respectively. We then combined the VLASS sources of each cluster and cross-matched with the combined MeerKAT sources of A2631 and ZwCL2341. We found a total of 798 MeerKAT/VLASS sources in A2631 and ZwCL2341 after a 5 arcsec cross-match. To minimize the positional accuracy uncertainties brought about by extended sources we select only compact sources with $0.8 < S_T/S_P < 1.2$. This criterion decreased the total number of MeerKAT/VLASS sources to 305.

We show the offsets in RA and DEC in Fig. 9 for the VLASS/MeerKAT combined cross-matched sources of A2631 and ZwCL2341. We have plotted $[1, 3] \times \sigma$, in red dotted lines, where $\sigma$ is standard deviation of offset. From Fig. 9, the offset of all sources lie within $3\sigma$ (<3 arcsec difference). Majority of sources are within $1\sigma$ (<1 arcsec difference). The mean offsets for RA and Dec. are $\Delta\alpha, \Delta\delta = 0.70$ arcsec, 0.32 arcsec, respectively and are shown by the horizontal dotted black lines. Given the differences in beam sizes of VLASS/MeerKAT the positional offsets of MeerKAT are therefore consistent with the VLASS reference positions and no offset to the RA, Dec. of the MeerKAT sources was applied.

### 4.4 Resolved versus unresolved sources

To assess which radio sources in our catalogues are resolved (sources with sizes larger than the synthesized beam) and which are unresolved (source size smaller than the synthesized beam) we make use the fact that the ratio of the total flux density to the peak flux density is a direct measure of the extension of a source:

$$S_{\rm total}/S_{\rm peak} = \theta_{\rm maj}\theta_{\rm min}/b_{\rm min}b_{\rm maj} \quad , \tag{1}$$

where $\theta_{\rm maj}\theta_{\rm min}$, $b_{\rm min}b_{\rm maj}$ are the source and synthesized beam FWHM axes, respectively. The total flux densities were computed using PyBDSF as discussed in Section 3.2. For a perfect gaussian unresolved source in the absence of noise, the peak brightness equals the total flux density $S_T/S_P = 1$. Unresolved sources with $S_T/S_P < 1$ or $S_T/S_P > 1$ result because of statistical errors in flux density error measurements and background noise.

We have used the method of M. Bondi et al. (2008) to determine the population of resolved and unresolved sources. Their method involves fitting a lower envelope to the $S_T/S_P$ ratio distribution such that it contains 95 per cent of all sources with $S_T < S_P$. This lower envelope curve takes the form of:

$$S_{\rm total}/S_{\rm peak} = A + \frac{B}{(S_{\rm peak}/\sigma_{\rm local})} \quad . \tag{2}$$

The values of $A$ and $B$ are obtained from fitting the data to contain 95 per cent of sources with $S_T < S_P$. For our data we found $(A, B) = (1.03, 1.01)$ and $(A, B) = (1.40, 1.70)$ for A2631 and ZwCL2341, respectively. After fitting the curve of equation (2) we plotted its mirror image by reflecting it about the $S_t/S_p = 1$ line. The resulting envelope functions are shown in Fig. 10 on a plot showing the ratio of the total ($S_T$) and peak ($S_P$) flux density as a function of their S/N ($S_P/\sigma$). Sources that lie between these two envelope curves are shown in green and classified as unresolved. All other sources outside these curves are classified as resolved. Unresolved sources had their total flux density set to the peak flux density ($S_T = S_P$) and their deconvolved angular size set to 0. Resolved sources had their total flux density and angular size unchanged. The percentage of unresolved/resolved sources determined from the fitted curves of each cluster is shown in Table 5.

## 5 SPECTRAL INDICES

The value of the spectral index can give us insight on the various particle acceleration mechanisms behind the powering a radio source. At GHz frequencies non-thermal synchrotron emission from AGN lobe-dominated sources and SFGs have a steep spectrum with $\alpha \sim -0.6$ to $-0.8$, while free–free emission from star-forming H II regions are much flatter with $\alpha \sim -0.3$ to $-0.5$ (J. J. Condon & S. M. Ransom 2016).

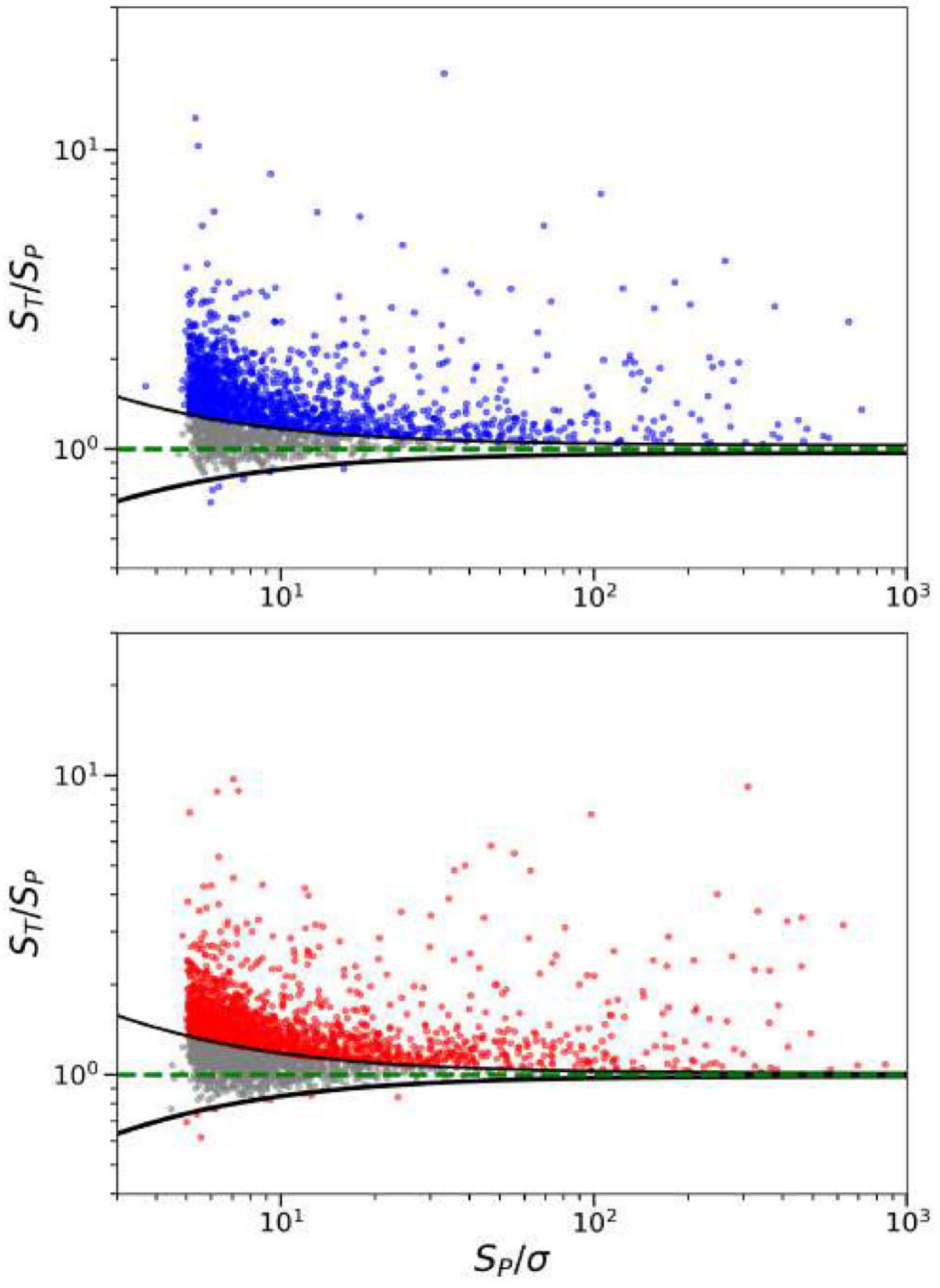

**Figure 10.** Ratio of the total flux density $S_\mathrm{T}$ to peak flux density $S_\mathrm{P}$ as a function of S/N ($S_\mathrm{P}/\sigma$) for catalogued sources in A2631 (top) and ZwCL2341 (bottom). The horizontal dotted line indicated the $S_\mathrm{T} = S_\mathrm{P} = 1$ line. Sources below this line have $S_\mathrm{P} > S_\mathrm{T}$ while sources above it have $S_\mathrm{T} > S_\mathrm{P}$. The points between the two solid curves show the unresolved source population. All sources outside these curves are classified as resolved.

**Table 5.** Derived parameters of equation (2) that determine the upper and lower envelope curves of Fig. 10 for finding the resolved/unresolved fraction of sources.

| Field | A | B | $\%_\mathrm{resolved}$ | $\%_\mathrm{unresolved}$ |
|---|---|---|---|---|
| A2631 | 1.03 | 1.40 | 69 | 31 |
| ZwCL2341 | 1.01 | 1.70 | 59 | 41 |

### 5.1 Spectral index distribution

For a source observed at frequency $\nu_1$, $\nu_2$ with flux density $S_1$, $S_2$ the spectral index $\alpha$ can be found from:

$$\alpha = \log\left(\frac{S_1}{S_2}\right) \Big/ \log\left(\frac{\nu_1}{\nu_2}\right) \quad . \tag{3}$$

For sources observed over multiple frequencies we need to fit the simple power law, $S_\nu \propto \nu^\alpha$ across all fluxes and frequencies. We have used three DDFACET subband images to derive the spectral index distribution between 998 and 1569 MHz. These subband images were created during the final DDFACET imaging run (see Section 2.1) using the option `Freq-NBand = 3`. We ran PYBDSF using the same source finding parameters as used for producing the main catalogue (see Section 3.2) on these three subband images to produce three catalogues at 998, 1283, and 1569 MHz. We then cross-matched between these catalogues resulting in a single catalogue with the same properties as main catalogue but at 998, 1283, and 1569 MHz. The resulting cross-matched catalogue contained 683 and 977 sources for A2631 and ZwCL2341, respectively.

For each source we fitted a simple power across 998, 1283, and 1569 MHz and measured their flux density in each corresponding image to find $\alpha$. We show the $\alpha$ distribution in Fig. 11 for A2631 (blue) and ZwCL2341 (red). The spectral index distribution for both clusters is well described by a single Gaussian with a population centred around $\alpha \approx -0.7$, which is in agreement from previous radio continuum studies and typical of synchrotron dominated emission (U. Lisenfeld & H. Völk 1999; A. E. Kimball & Ž. Ivezić 2008; V. Smolčić et al. 2017).

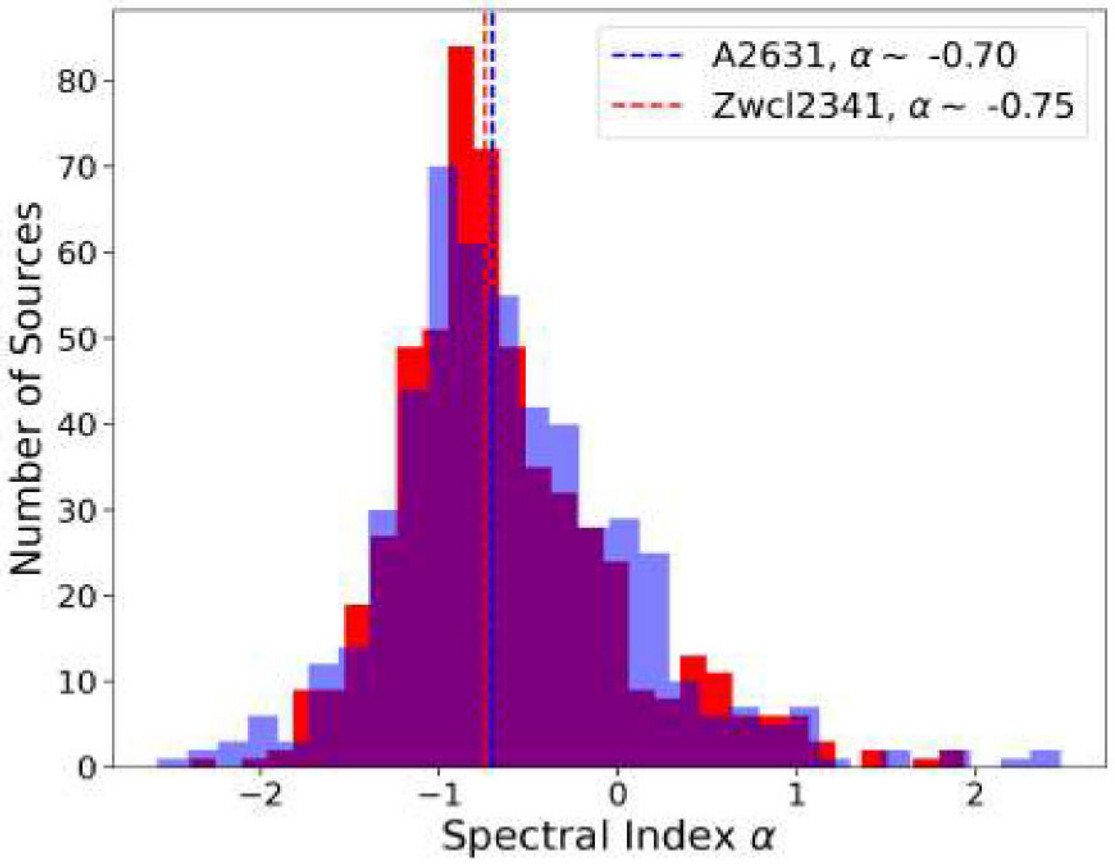

**Figure 11.** Spectral index distribution for sources between 998–1569 MHz for A2631 and ZwCL2341. The red and blue dotted vertical line shows the respective median spectral index for the entire population.

### 5.2 Spectral index flux density relation

The change in spectral index as a function of flux density corresponds to a gradual change in the underlying physical processes of the different galaxy populations. We have plotted in the left side of Fig. 12 the combined spectral indices of A2631 and ZwCL2341 between 998 and 1596 MHz (blue and red in Fig. 11) versus their 1283 MHz flux density. We then plotted the median spectral index value in equally spaced logarithmic bins shown by the black boxes. There is a clear trend of decreasing spectral index value from very steep $\alpha = -1$ at the high-flux end towards very flat $\alpha = -0.3$ at the faint flux end. This trend of spectral index flattening towards fainter flux densities could be explained by the splitting of the radio source population into two distinct populations of steep and flat spectrum sources.

To understand the individual radio source populations giving rise to this trend we made use of the radio luminosity. In the local universe the radio luminosity function is dominated by AGN-related emission above $L_{1.4\mathrm{GHz}} \geq 10^{24}$ W Hz$^{-1}$, while radio emission from star-forming processes in galaxies dominates below this threshold (J. Condon, W. Cotton & J. Broderick 2002; A. E. Kimball et al. 2011b; J. Condon et al. 2013). We therefore converted the radio flux densities at $\nu_\mathrm{obs} = 1.28$ GHz to the 1.4 GHz radio

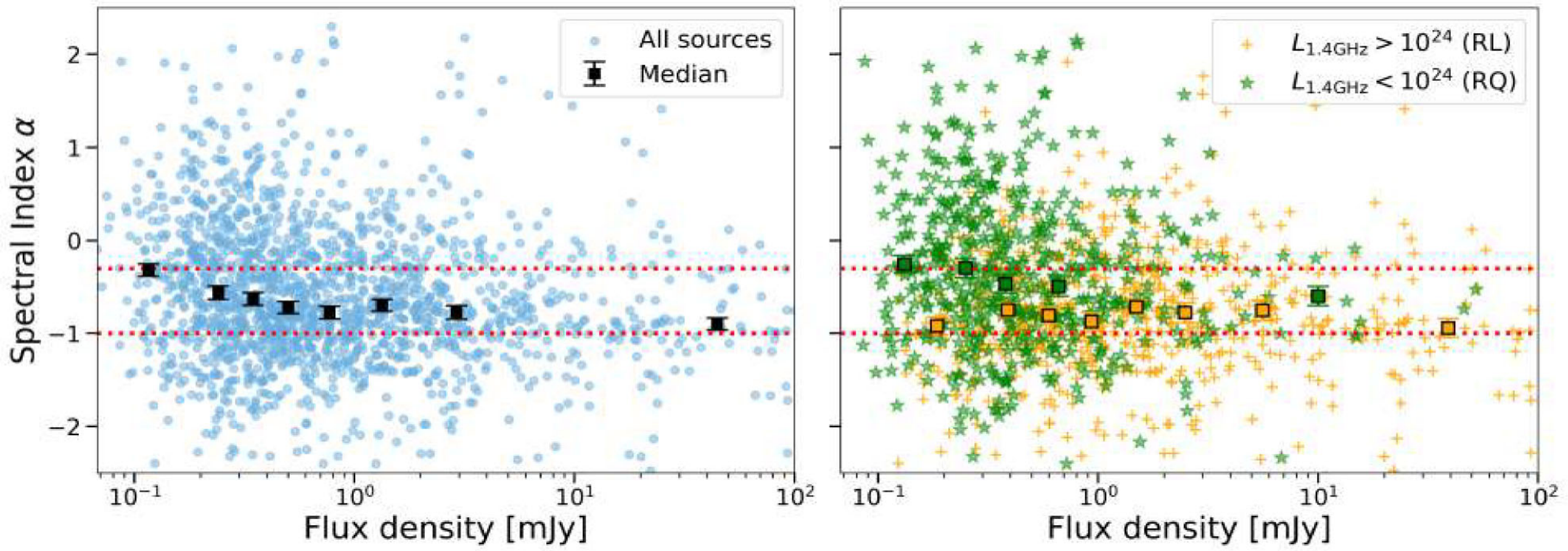

**Figure 12.** Left: the combined spectral index distribution of all sources in A2631 and ZwCL2341 between 998 and 1569 MHz versus their flux density at 1283 MHz. The horizontal lines at $\alpha = -0.3$ and $\alpha = -1$ are extremes for typical values of very flat and very steep spectrum sources. The black points show the median spectral index values in equally spaced logarithmic bins with errors. There is a clear trend of flattening in the spectral index distribution towards fainter flux densities. Right: the plot on the left split into into two different groups of RL sources defined by $L_{1.4\mathrm{GHz}} > 10^{24}$ (stars) and RQ sources defined by $L_{1.4\mathrm{GHz}} < 10^{24}$ (crosses). RQ sources maintain a relatively flat spectral index range while for RL sources, the spectral index is relatively steep.

luminosity $L_{1.4\mathrm{GHz}}$ at the cluster redshift $z$:

$$L_{1.4\mathrm{GHz}} = \frac{4\pi D_{\mathrm{L}}^2(z)}{(1+z)^{1+\alpha}} \left(\frac{1.4\mathrm{GHz}}{\nu_{\mathrm{obs}}}\right)^{\alpha} S_{\nu\mathrm{obs}} \quad . \tag{4}$$

Where $D_{\mathrm{L}}$ is the luminosity distance and $\alpha$ is the derived spectral index from Section 5.1.

We used photometric redshifts from Hyper Supreme Cam (HSC) WIDE survey to obtain $z$ for calculation of $L_{1.4\mathrm{GHz}}$ in equation (4) for the galaxies that have spectral index values. We used the MIZUKI[7] (M. Tanaka 2015; M. Tanaka et al. 2018) catalogue from HSC WIDE to obtain the photometric redshifts. We chose to use the `photoz_best` column for our redshift estimates and selected only those with a risk factor[8] <0.5 to ensure reliable photometric redshift estimates. The HSC catalogues were cross-matched with our radio catalogues using a radius of 3 arcsec (see Section B for justification). 95 per cent of the MeerKAT radio sources had a corresponding HSC WIDE counterpart after cross-match.

We split our sample into two distinct groups using a cut in the radio luminosity: RL sources characterized by $L_{1.4\mathrm{GHz}} > 10^{24}$ and RQ sources characterized by $L_{1.4\mathrm{GHz}} < 10^{24}$ W Hz$^{-1}$. We show on the left side of Fig. 12 the spectral index values versus flux density for these RL and RQ sources in orange and green, respectively. We see that there is a clear separation between the spectral index values of RL and RQ sources. RL are generally characterized by a steeper spectrum as compared to RQ sources. RQ sources mostly occupy the fainter flux density region (with majority <2mJy) and maintain an overall flat spectral index value between $-0.3 < \alpha < -0.5$. The RL occupy a range of flux densities (0.1–100 mJy) and maintain an overall steep spectral index value between $-0.6 < \alpha < -0.9$.

To understand the trends of these two samples we need to understand their respective dominate emission mechanisms attributed to each. Star formation processes associated with SFGs are expected to dominate at $L_{1.4\mathrm{GHz}} < 10^{24}$ W Hz $^{-1}$ (J. Condon 1992). Since both synchrotron and thermal bremsstrahling (free–free) emission processes are attributed to SFGs, we would expect a mixture of steep and flat spectrum sources as observed in our plot. However, since free–free emission dominates at higher frequencies compared to synchrotron-related processes (J. Condon 1992; M. Clemens et al. 2008), we expect our SFGs sample to have an overall flatter spectrum. From Fig. 12, the clear trend of flat spectrum RQ sources are most likely attributed to SF processes (I. Prandoni et al. 2006; A. E. Kimball et al. 2011a; J. Condon et al. 2013; I. Whittam et al. 2013; K. Kellermann et al. 2016). Steep spectrum RL sources are most likely synchrotron powered by AGN emission associated with large relativistic jets and lobes (T. M. Heckman & P. N. Best 2014; K. Kellermann et al. 2016). Together the steep spectrum RL and flat spectrum RQ sources explain the spectral index flattening towards fainter flux densities trend observed in the left of Fig. 12.

## 6 RADIO SOURCE COUNT CORRECTIONS

In deriving the true underlying source count population one must first take into account the various selection effects causing incompleteness in our radio survey. The preferential non-detection of large resolved sources is termed resolution bias, the preferential detection of unresolved sources close to detection threshold is termed Eddington bias. The inhomogeneity of the noise across the FoV also causes preferential non-detection of faint sources. These biases can either be estimated all at once through sophisticated simulations with realistic flux density and source size distributions or estimated individually. We have chosen to use the latter approach. We have used simple simulations to account only for the noise bias and then used the appropriate relations to account resolution and Eddington bias.

[7]MIZUKI is a template SED fitting code that uses Bayesion priors on physical properties of galaxies to compute the photometric redshift.
[8]The risk factor is the reliability of the estimated photometric redshift that has a value between 0 and 1, with 0 being the least risk and hence most reliable. A value of 0.5 is moderate risk.

### 6.1 Noise bias and radio simulations

The noise of radio images is non-uniform and composed of regions of varying RMS, as seen from Fig. 6. This means that faint sources found in regions of relatively high RMS (noise peaks) can have their flux density dip below the detection threshold of the source finder much easier as compared to faint sources found in relatively low RMS regions, resulting in the source finder being unable to detect them. This is known as noise bias and is mainly an issue for faint sources (I. Prandoni et al. 2000). To account for this noise effect on the completeness we make use of Monte Carlo simulations.

#### *6.1.1 Injection and retrieval of mock sources*

We injected 500 mock sources at random positions in 50 equally spaced bins in a total flux density range of 0.03–10 mJy. The lower limit of 0.03 mJy was chosen since the faintest flux density in both catalogues does not fall below this value. Simulations were performed separately for each bin, that is, 500 sources were injected on separate images for each of the 50 bins to avoid the problem of blending. The images used for injection of sources where the residual maps produced from PyBDSF (`img_type = 'gaus_resid'`). This is the real image (right images of Fig. 3) with the modeled Gaussian components subtracted. Sources are then recovered on the resulting image (injected + residual) using the same source finding properties used for creation of the real catalogue (see Section 3.2).

#### *6.1.2 Flux density and source size distribution*

For each bin source fluxes were randomly chosen between the left and right bin edges. For the source sizes, The FWHM major axis was randomly chosen between 1 and 2 times the synthesized beam size major axis $b_{\mathrm{maj}}$, while the FWHM minor axis was chosen between 0.5 and 1 times FWHM major axis. The positional angle was randomly chosen between 0 and 180 deg. For a realistic representation of sources we injected the same fraction of unresolved and resolved sources as was found in the real catalogue (see Section 4.4).

#### *6.1.3 Simulation caveats*

A constraint on the minimum distance between two sources was implemented. Sources were not injected at distances less than 2 times the beam size (≈20 arcsec) between other sources. It must be noted that spurious sources due to residual calibration errors exist at $> 5\sigma$ in the residual maps. To remove the contribution of these sources, we ran the source finder on the residual image using the same source finding properties as used on the real image. We retrieved a total of 106 sources for A2631 and 205 sources for ZwCL2341. This number was subtracted from the total number of recovered sources for each simulation.

#### *6.1.4 Noise bias correction factors*

The detected fraction was found by dividing the number of recovered sources over the total number of injected sources (500) for each bin. We show in Table 6 the noise bias correction factors and error. We show in Fig. 13 the resulting plot of Table 6 (detected fraction). The error was taken to be the Poisson error of the injected sources with recovered sources added in quadrature. We see that the correction factor reaches a completeness of 1 for fluxes >1 mJy.

**Table 6.** Completeness for the MeerKAT 1.28 GHz A2631 (top) and ZwCL2341 (bottom) catalogue as a function of flux density.

| Flux density (mJy) | Completeness ($C_{\mathrm{comp}}$) | Error |
|---|---|---|
| 0.0319 | 0.002 65 | 0.0476 |
| 0.0358 | 0.003 55 | 0.0323 |
| 0.0403 | 0.009 57 | 0.0114 |
| 0.0452 | 0.0174 | 0.006 06 |
| 0.0509 | 0.0318 | 0.003 25 |
| 0.0572 | 0.0518 | 0.001 97 |
| 0.0642 | 0.0916 | 0.001 11 |
| 0.0722 | 0.136 | 0.000 743 |
| 0.0812 | 0.194 | 0.000 519 |
| 0.0912 | 0.262 | 0.000 384 |
| 0.102 | 0.333 | 0.000 302 |
| 0.115 | 0.408 | 0.000 246 |
| 0.130 | 0.488 | 0.000 205 |
| 0.146 | 0.562 | 0.000 178 |
| ⋮ | ⋮ | ⋮ |
| 0.945 | 0.992 | 0.000100 |
| 0.0319 | 0.006 56 | 0.0189 |
| 0.0358 | 0.0124 | 0.009 35 |
| 0.0403 | 0.0212 | 0.005 26 |
| 0.0452 | 0.0367 | 0.002 98 |
| 0.0509 | 0.0636 | 0.001 69 |
| 0.0572 | 0.101 | 0.001 05 |
| 0.0642 | 0.147 | 0.000 715 |
| 0.0722 | 0.203 | 0.000 516 |
| 0.0812 | 0.259 | 0.000 400 |
| 0.0912 | 0.326 | 0.000 316 |
| 0.102 | 0.385 | 0.000 267 |
| 0.115 | 0.460 | 0.000 222 |
| 0.130 | 0.525 | 0.000 193 |
| ⋮ | ⋮ | ⋮ |
| 0.945 | 0.976 | 0.000 103 |

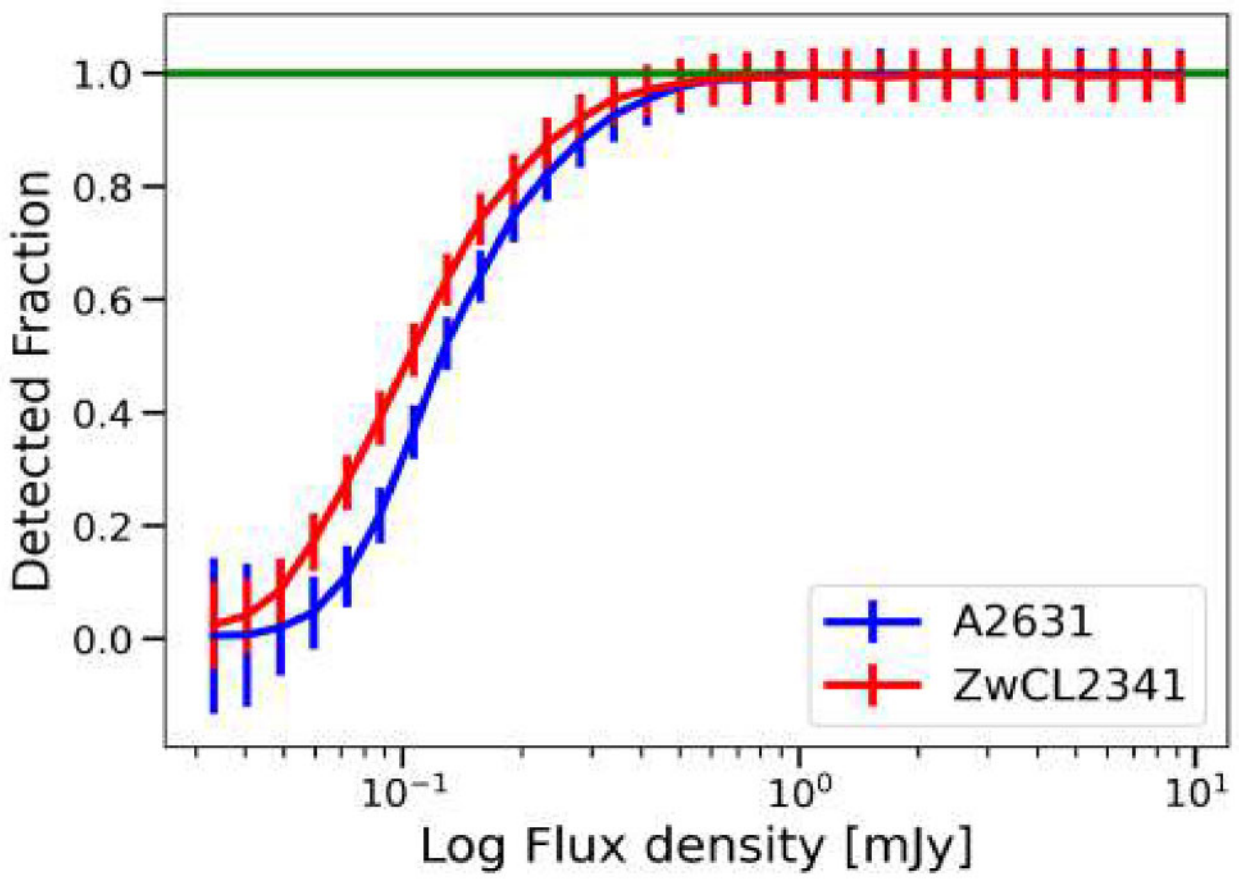

**Figure 13.** Fraction (solid line) of recovered over injected sources for a total of 20 simulations for A2631 and ZwCL2341. The horizontal green line represents a fraction = 1 (number of injected sources equal to recovered sources).

### 6.2 Size distribution and resolution bias

The source size distribution of the resolved source population of a particular field will vary from observation and catalogue source finding parameters. This induces a bias in the source sizes that is directly related to the total flux density as seen from equation (1). A source of a given total flux density but larger angular size will have the total flux density spread over a larger area resulting in a lower peak flux density compared to a source of smaller angular size with same total flux density, thereby making it easier to fall below the detection threshold of a survey. This effect is known as resolution bias and is needed if we wish to obtain an unbiased consensus on the resolved source population. From I. Prandoni et al. (2000, 2006), we can derive the maximum angular size $\Theta_{\rm max}$ in which a source can have before its peak drops below the detection threshold ($5\sigma$).

In a survey with a fixed resolution, sources with the same total flux density but larger angular sizes will have lower peak flux densities because their emission is spread over more beams

$$\Theta_{\rm max} = \Theta_{\rm N}\sqrt{((S_{\rm total}/5\sigma)) - 1} \quad , \tag{5}$$

where $\Theta_{\rm N} = \sqrt{b_{\rm maj}b_{\rm min}}$ is the geometric mean of the beam FWHM. In addition to $\Theta_{\rm max}$, there is also a minimum angular size $\Theta_{\rm min}$ that takes into account the finite nature of the synthesized beam and ensures that the limiting size does not become unphysical ($\theta_{\rm max} \to 0$ at low S/Ns). $\Theta_{\rm min}$ can be obtained by combining equation (1) with equation (2):

$$\Theta_{\rm min} = \Theta_{\rm N}\sqrt{\alpha + \beta/(S_{\rm total}/\sigma) - 1} \quad . \tag{6}$$

Sources with sizes $> \Theta_{\rm max}$ will remain undetected contributing to the incompleteness of the catalogue while for sources with sizes $<\Theta_{\rm min}$ the deconvolution is not reliable and must therefore be excluded. It follows from I. Prandoni et al. (2001) that the resolution bias is defined as the maximum of these two sizes:

$$\Theta_{\rm lim} = \max(\Theta_{\rm max}, \Theta_{\rm min}) \quad . \tag{7}$$

In Fig. 14, we have plotted the deconvolved sizes ($\Theta$) as a function of total flux density ($S_{\rm T}$). Due to their identical beam size and similar sensitivities we have chosen to combine the deconvolved sizes of A2631 and ZwCL2341 in this plot. The deconvolved sizes were defined to be the geometric mean of the FWHM major and minor axis. For unresolved sources (see Fig. 10), we have set their sizes to be $\Theta = 0$. The range of possible values for $\Theta_{\rm min}$ and $\Theta_{\rm max}$ depends on the RMS levels and are indicted by the dashed orange and green curves, respectively. To define these RMS levels we consider the maximum and minimum noise values in our map. There is a central RMS of $\sigma \sim 16, 11$ µJy for A2631 and ZwCL2341, respectively. We therefore chose $\sigma \sim 16$ µJy to represent the minimum noise value. For the maximum noise value, this usually represents the noise at the edges. We can see from Fig. 7, about 99 per cent of all pixels have an RMS value less than ∼60 µJy for both A2631 and ZwCL2341. This value was therefore taken to represent the maximum noise value.

In Fig. 14, we have plotted the maximum and minimum allowed size $\Theta_{\rm max}$, $\Theta_{\rm min}$ given by equations (5) and (6) in dashed green and orange, respectively. There are two curves for each $\Theta_{\rm max}$, $\Theta_{\rm min}$, one for the minium noise and one for the maximum noise value. The two curves shift to the right with increasing noise value. We see that the source sizes all tend to be smaller than $\Theta_{\rm max}$ and larger than $\Theta_{\rm min}$, we also see in most cases $\Theta_{\rm lim} = \Theta_{\rm max}$, except for the fainter flux densities below 0.1 mJy, where $\Theta_{\rm lim} = \Theta_{\rm min}$ for the maximum noise curves.

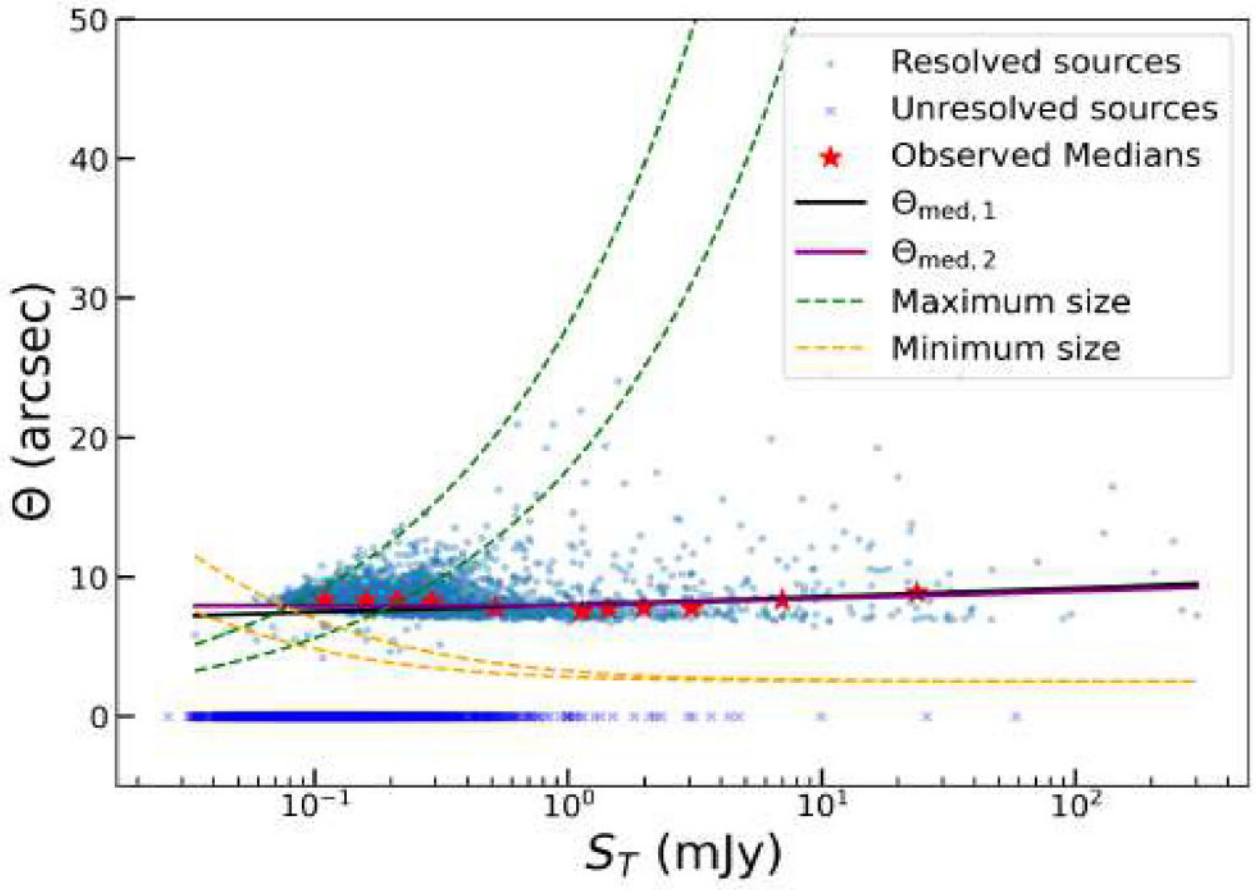

**Figure 14.** Angular size $\Theta$ (geometric mean of FWHM major/minor axis) as a function of their total flux density for the combined sources of A2631 and ZwCL2341. The upper and lwoer dashed curves represent the the maximum and minimum allowed angular sizes $\Theta_{\rm max}$ and $\Theta_{\rm min}$, respectively. The leftmost curves of $\Theta_{\rm min}$ and $\Theta_{\rm max}$ were derived using a central RMS of $\theta_{\rm c} = 16$ µJy. The right-shifted curve of the latter $\Theta_{\rm min}$ and $\Theta_{\rm max}$ was derived using an RMS of 60 µJy (associated with the edge region). Sources above $\Theta_{\rm max}$ and below $\Theta_{\rm min}$ are affected by resolution bias. All sources classified as unresolved using the method in Section 4.4 have their angular sizes set to 0. The median source sizes calculated from equally spaced logarithmic bins are shown as red stars.

In Fig. 14, we have also plotted the median angular sizes (red stars) using modified Windhorst relations of intrinsic angular sizes in an attempt to model the observed median source sizes in our catalogues. The Windhorst relations (R. Windhorst et al. 1990a) are used extensively in literature to estimate the resolution bias for 1.4 GHz surveys. They are defined as $\Theta = k(S_{\rm 1.4GHz})^m$, with $k = 2$ arcsec and $m = 0.3$. Our modified version of the Windhorst relations are:

$$\Theta_{\rm med,1} = 4 \times k\ (S_{\rm 1.4GHz})^{m\times 0.1}\ \text{arcsec},$$
$$\Theta_{\rm med,2} = \begin{cases} (4\times k)\ (S_{\rm 1.4GHz})^{m\times 0.1}\ \text{arcsec} & S_{\rm 1.4GHz} > 1\,\text{mJy} \\ (4\times k)\,\text{arcsec} & S_{\rm 1.4GHz} < 1\,\text{mJy}, \end{cases}$$

We note that $4 \times k$ is of the order of the $b_{\rm maj}$. We see in Fig. 14 that $\Theta_{\rm med,1}$ follows the observed median sizes well for $S_{\rm T} > 1$mJy but diverges slightly for $S_{\rm T} < 1$mJy. We have therefore decided to use the modification of $\Theta_{\rm med,2}$ with constant median size for $S_{\rm T} < 1$mJy for $S_{\rm T} > 1$mJy much better describes our observed median source sizes over the entire flux density range. We therefore employ the $\Theta_{\rm med,2}$ relation to account for the resolution bias in our catalogues.

Knowledge of the true angular size distribution for these MeerKAT sources is crucial if we wish to account for the fraction of sources larger than $\Theta_{\rm max}$ and therefore the resolution bias. R. Windhorst, D. Mathis & L. Neuschaefer (1990b) reports an empirical integral distribution as a function of flux density in the form of:

$$h(>\Theta_{\rm lim}) = \exp\left[b\left(\frac{\Theta_{\rm lim}}{\Theta_{\rm med}}\right)^{a}\right]. \tag{8}$$

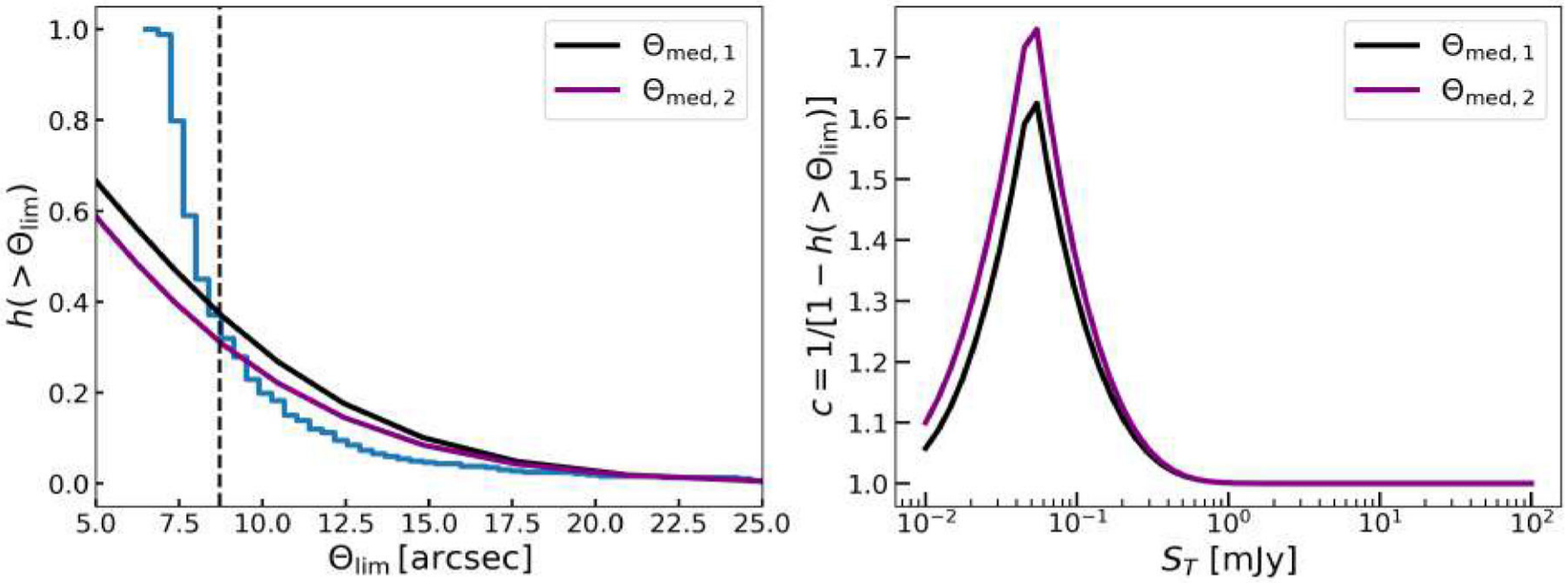

**Figure 15.** Left: source size cumulative distribution for the combined sources of A2631 and ZwCl2341 final catalogues shown together with the true integral size distribution of equation (8) for the median source sizes of $\Theta_{med,1}$ and $\Theta_{med,2}$ (equation 6.2). The sizes chosen for the integral sizes distribution are those unaffected by resolution bias in the regions bounded by shaded curves in Fig. 14 in the range $100 > S_T > 1$mJy. The vertical black dashed line is an approximate indication of the minimum angular size of which the observed distribution can be considered reliable. Right: the corresponding resolution bias for $\Theta_{med,1}$ and $\Theta_{med,2}$ calculated from equation (9).

With $a = -ln(2)$ and $b = 0.62$. To determine an unbiased integral size distribution we need to select sources in a total flux density that are not affected by the resolution bias (sources bounded by $\Theta_{min}$ and $\Theta_{max}$ curves). For this purpose, we choose the flux density range of $1 < S_T < 100$ mJy. From Fig. 14, it can be seen that this flux density range excludes most sources greater than $\Theta_{max}$ and less than $\Theta_{min}$ for the upper limit noise value.

We show on the left of Fig. 15 the real (observed) integral size distributions for the resolved sources (blue circles) in Fig. 14 for the range $1 < S_T < 100$ mJy. We also plot the true (theoretical) integral size distributions from equation (8) as a function of $\Theta_{lim}$ for the median source size relations of $\Theta_{med,1}$ (black curve) and $\Theta_{med,2}$ (purple curve). For both $\Theta_{med,1}$ and $\Theta_{med,2}$, we have fitted the parameters $a$, $b$ from equation (8) to the observed sizes. We have found $a, b = -0.97, 1.64$ and $a, b = -1.13, 1.45$ for $\Theta_{med,1}$ and $\Theta_{med,2}$, respectively. We see that the true integral size distribution for both median source size models provides a good fit for the observed sizes $> \sim 8$ arcsec (vertical dashed black line)[9]. However, $\Theta_{med,2}$ follows the observed integral size distributions slightly better than $\Theta_{med,1}$ and hence why we use it to account for the resolution bias.

### *6.2.1 Correction for resolution bias*

The correction factor $c$ that needs to be applied to the source counts to account for the resolution bias is defined from I. Prandoni et al. (2001) as:

$$c = 1/[1 - h(>\Theta_{lim})], \quad (9)$$

Where $h(>\Theta_{lim})$ takes the form of the integral of the angular size distribution of R. Windhorst et al. (1990a) of equation (8) and $\Theta_{lim}$ is the limiting angular size, above which the catalogues is expected to become incomplete. We show on the right of Fig. 15 the correction factor $c$ corresponding to $\Theta_{med,1}$ and $\Theta_{med,2}$. For the bright ($S_T > 1$ mJy) and faint ($S_T < 0.01$ mJy) sources the correction factor is negligible ($c \sim 1$). Sources with fluxes between $0.02 < S_T < 0.2$ mJy, the correction factors are largest with values ranging between $1.2 < c < 1.8$.

[9] Below this limit the convolution is unreliable and sources sizes should be set to $\Theta = 0$.

### 6.3 Eddington bias

Eddington bias (A. Eddington 1913, 1940) takes into account the flux boosting that occurs mostly for faint sources close to the detection threshold. Due to random error measurements, the measured flux density values will be redistributed around their true value. Since the radio source population follows a non-uniform flux density distribution with increase in source population for deceasing flux density, the flux density shifting therefore disproportionally affects faint source population.

Faint sources with true flux density values close or slightly below the detection threshold that have their flux density boosted will be detected as a result. Bright sources boosted above or below their true value will be detected nevertheless. Therefore the probability to detect a source below the detection threshold is higher than the probability to miss a source above the detection threshold. The completeness of the catalogue is therefore biased at the detection threshold. We have used the maximum likelihood approach of C. Hales et al. (2014) to correct the fluxes to their true Eddington unbiased value

$$S_{true} = \frac{S_{obs}}{2}\left(1 + \sqrt{1 - \frac{4\gamma}{(S/N)^2}}\right) \quad , \quad (10)$$

where $\gamma = \gamma(S)$ is the slope of the counts at a given flux density $dN/dS \sim S^{-\gamma}$, and S/N is the signal to noise ratio. The slope can be derived from a polynomial fit of the observed counts and is given by:

$$\gamma = 2.5 - \sum_{i=0}^{n} i \cdot a_i (\log S)^{i-1} \quad . \quad (11)$$

In deriving $\gamma$ we used the polynomial form of the 1.4 GHz source counts available in literature. We chose to compare the commonly used M. Bondi et al. (2008) 1.4 GHz fitted model with the newer semi-empirical models of T-RECS (A. Bonaldi et al. 2023)[10] and SEMPER (M. Giulietti et al. 2025). Since (M. Giulietti et al. 2025) only deals with the SFG population we combined it with the AGN model of C. Mancuso et al. (2017).[11] The polynomial fit for M. Bondi et al. (2008) has been previously derived equation (2) whereas for M. Giulietti et al. (2025) and A. Bonaldi et al. (2023) we derived it from the available data. The results of these models in deriving the slope $\gamma$ of equation (11) are shown in the top of Fig. 17.

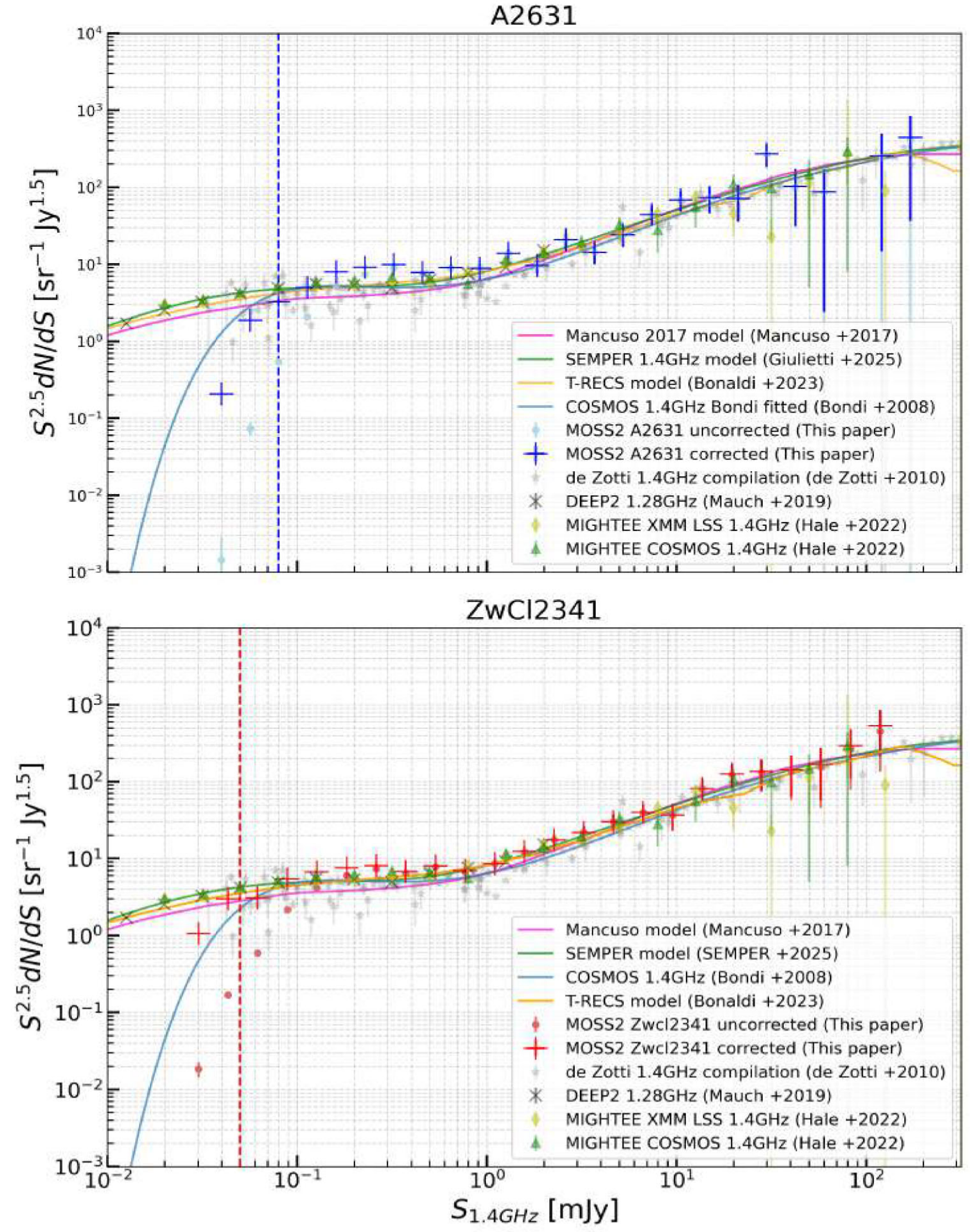

**Figure 16.** Normalized corrected and uncorrected 1.4 GHz differential source counts for A2631 (blue) and ZwCL2341 (red). The corrected counts are indicated by the large crosses for A2631 and ZwCL2341. The uncorrected counts are represented by the faint circles. Error bars correspond to Poisson counting errors. The solid curves indicate the 1.4 GHz AGN + SFG models of C. Mancuso et al. (2017), SEMPER (M. Giulietti et al. 2025), TRECS (A. Bonaldi et al. 2023) and M. Bondi et al. (2008), respectively. The faint stars represent a compilation of 1.4 GHz source counts from G. De Zotti et al. (2010). The crosses, diamonds, and triangles represent other deep MeerKAT data in the form of DEEP2 (T. Mauch et al. 2020), MIGHTEE XMM LSS (C. Hale et al. 2023), and MIGHTEE COSMOS (C. Hale et al. 2023), respectively. The dashed vertical lines in each plot represent the detection limit for each cluster $5 \times \sigma$, where $\sigma$ is the central RMS for each cluster.

The slope $\gamma$ of the newer models of M. Giulietti et al. (2025) and A. Bonaldi et al. (2023) are generally consistent across a wide flux density range. They are also somewhat comparable to M. Bondi et al. (2008) at high-flux density ∼100 mJy, whereas at the very faint end <0.1 mJy, the discrepancy is the largest. At the faint end, the slope of the M. Bondi et al. (2008) model shows a rapid drop due to unreliable fits from sensitivity constraints unable to capture the sufficient faint SFG population. M. Bondi et al. (2008) reaches a maximum slope of $\gamma \sim 2.6$ before rapidly dropping whereas for M. Giulietti et al. (2025) and A. Bonaldi et al. (2023), the maximum $\gamma \sim 2.2$ is achieved at around $\sim$0.1 mJy before it then gradually drops. We show at the bottom of Fig. 17 the expected flux density boosting expected from these three different models for S/N of 5, 8, 20. For higher S/N sources the flux density boosting is negligible whereas for the lower S/N sources a flux density boost of around $\sim$10 per cent is expected. We see that $\gamma$ for M. Bondi et al. (2008) is not able to reach the 0.01 mJy regime where the detection threshold of our data resides and therefore results in non-physical flux density boosting ratios at these faint flux density limits.

[10]We used the'catalog_continuum_wrapped.fits' catalogue which includes both the AGN and SFG populations.
[11]The AGN counts are found here: http://w1.ira.inaf.it/rstools/srccnt/srccnt_tables.html

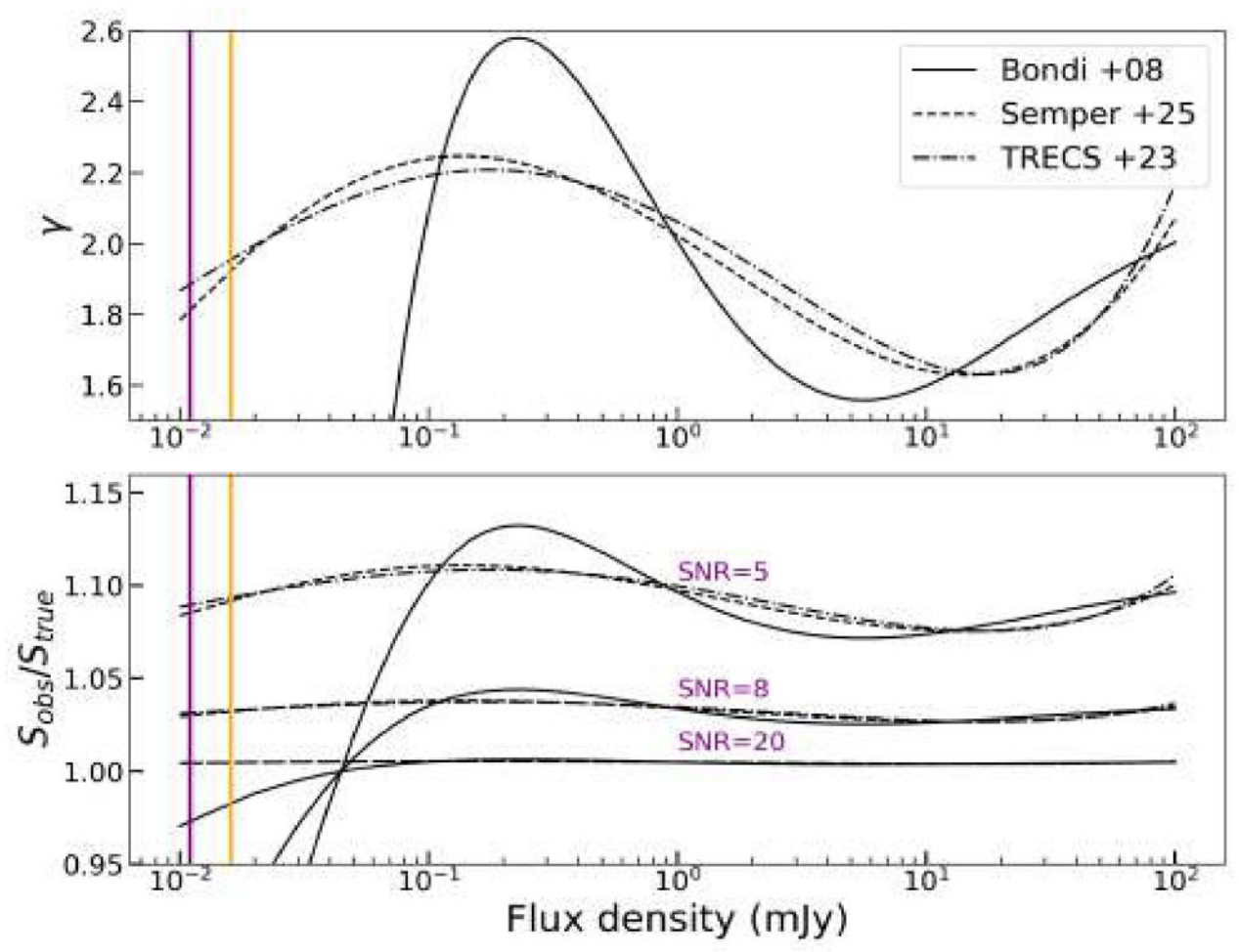

**Figure 17.** Eddington bias for different number count distributions models of M. Bondi et al. (2008) (solid line) with the newer semi-empirical models of A. Bonaldi et al. (2023) (dotted–dashed line) and M. Giulietti et al. (2025) (dashed line) at 1.4 GHz. The detection threshold of A2631 and ZwCl2341 is shown by the vertical lines, respectively. The top panel shows the slope of the source counts $\gamma$; $dN/dS \sim S^{\gamma}$ derived from the sixth order polynomial of these source count models. The bottom panel shows the flux boosting ratio observed for the measured flux densities ($S_{\mathrm{obs}}$) at different S/N ratios of $S/N = [5, 8, 20]$. Lower S/N sources experience the most significant flux boosting.

Our choice of model to correct for Eddington bias therefore lies between the newer models of A. Bonaldi et al. (2023) and M. Giulietti et al. (2025). Since both models are very similar we have chosen to use the newest model of M. Giulietti et al. (2025) to account for the Eddington bias in our data.

### 6.4 False detection rate

Residual calibration errors can manifest themselves as noise peaks in the image which can be incorrectly classified as real sources (false detentions). Since we cannot quantify these directly we can use the assumption that for very positive noise peak there exists a corresponding negative noise peak. We can therefore use these negative peaks as a means to indirectly quantify the number of false detections. This is done by inverting the real continuum map (multiply by $-1$) and running the PyBDSF using the same settings used for real catalogue on the inverted map. A total of 25 and 33 false detections were found for A2631 and ZwCL2341, respectively. 5 and 11 of which have $S/N > 10$ for A2631 and ZwCL2341, respectively. These false detections were mostly found in pairs of 2 or more (with 7 being the most for a single source) around bright sources exhibiting strong artifacts. For A2631 and ZwCL2341, the total fraction of spurious sources in our catalogue is therefore $\sim$1 per cent for both clusters making them negligible.

### 6.5 Multicomponent sources

The source finding procedure of PyBDSF often encounters difficulties in correctly fitting the correct number of Gaussian components to sources of complex morphologies. Single radio sources with multiple different components can be mistakenly classified as separate entries in the catalogue resulting in multiple entries for the same object. The total flux density of these types of sources need to be summed otherwise it could have a negative impact on the source counts calculation.

To identify these sources we used the flux-separation algorithm of M. Magliocchetti et al. (1998) together with the source classification information given by PyBDSF. The M. Magliocchetti et al. (1998) algorithm was used in FIRST to identify tight groups of sources likely to be subcomponents of a single source. This algorithm compares the separation and total flux density ratio between a source and its nearest neighbour. We have considered a component to be part of a another source if their flux density ratio $f$ is in the range of $0.25 \leq f \leq 4$, and satisfies the separation

$$\Theta < 100\sqrt{\frac{S_{\mathrm{T}}}{20}}\ , \tag{12}$$

where $\theta$ is in arcseconds and $S_{\mathrm{T}}$ is the summed flux density of the two components. PyBDSF, as described in Section 3.2 produces an output column `S_Code` in which a source is classified according to the number of Gaussians fitted and grouped into a single island. Due to their definition, sources with `S_Code` = 'C' or 'M' are most probable to be multicomponent source. We have therefore only chosen sources with `S_Code` = 'C' or 'M' to be used with this algorithm to determine our sample of potential multicomponent sources. A total of 329 and 313 sources of `S_Code` = 'C' or 'M' was found for A2631 and ZwCL2341, respectively. After evaluating these sources with the algorithm of M. Magliocchetti et al. (1998) we find 51 and 68 double sources for A2631 and ZwCL2341, respectively which is <3 per cent of the entire catalogue for each cluster.

The M. Magliocchetti et al. (1998) algorithm can sometimes incorrectly classify close-proximity individual sources as multicomponent. To investigate this, we obtained HSC (S. Miyazaki et al. 2012) Red Green Blue (RGB) colour optical cutouts and overlayed radio contours with their corresponding radio positions. We show in Fig. 18 a few of these cutouts. The green circles indicate the radio positions that were classified as a double source according to the M. Magliocchetti et al. (1998) algorithm. The white contours represent the 1.28 GHz radio emission. From these cutouts we see that some single-component sources are incorrectly classified as multicomponent as their radio detections each have clear optical counterparts. This is evident from the first two cutouts at the top. For the others, two or more radio detections with only a single or no optical counterpart exists suggesting these sources are in fact multicomponent in nature and have been correctly classified. In conclusion, the number of multicomponent sources are even smaller than we estimated making their uncorrected total flux density contribution in calculations involving the source counts negligible.

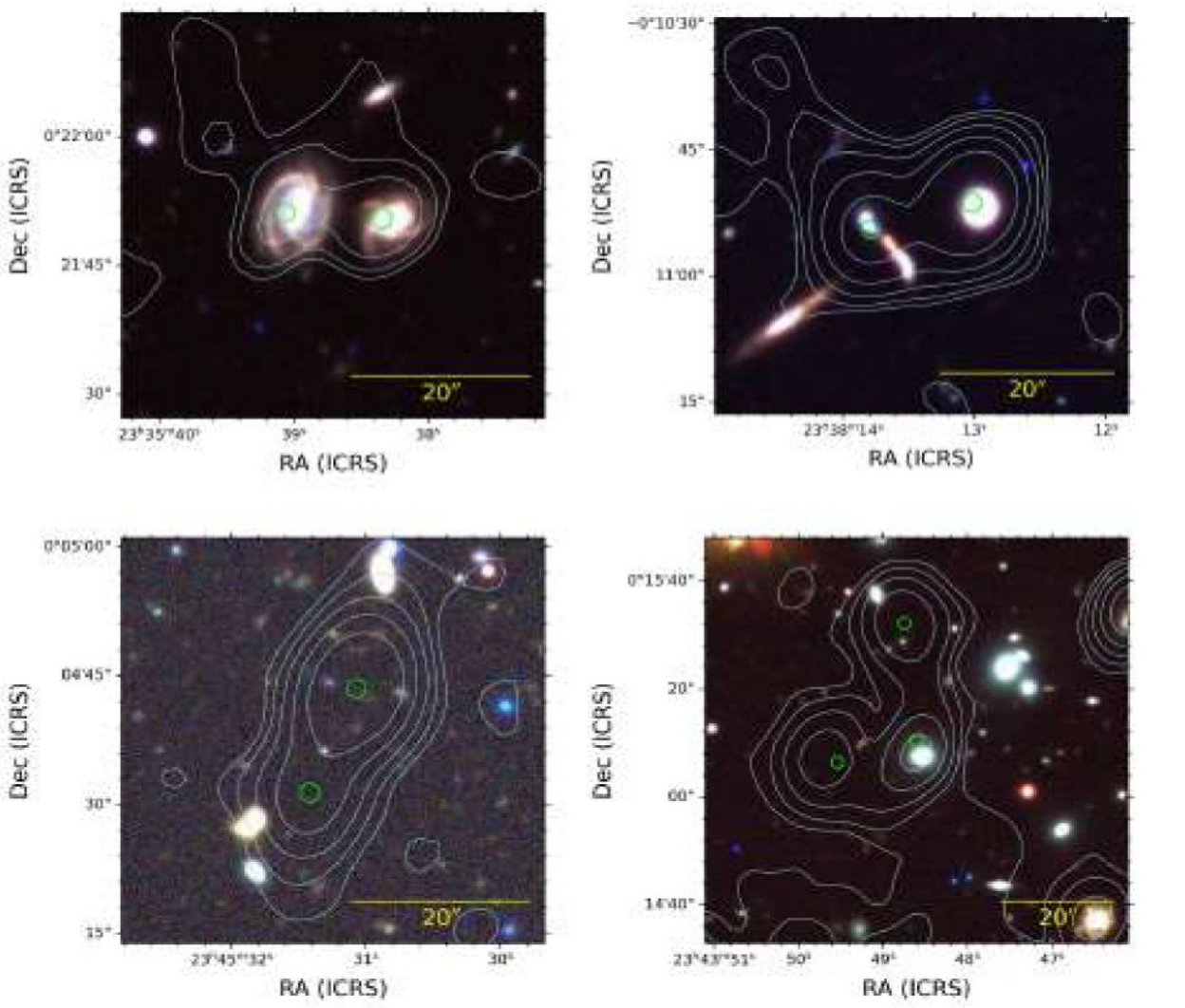

**Figure 18.** HSC Optical RGB images overlayed with radio contours (white) for the sources classified as multicomponent from the M. Magliocchetti et al. (1998) algorithm. The radio positions are shown as green circles. The radio contours are drawn for $2^i \times \sigma, i = [1, 2, 3, 4, 5]$, with $\sigma = 10, 15$ μJy for A2631 and ZwCL2341, respectively. We can see that some radio sources have been incorrectly classified as multicomponent due to the existence of each radio source having a clear optical counter part. This can be seen from the first two cutouts. For the others, due to their morphology and the non-existence of a corresponding optical counterpart, the radio sources are different components of a single system and are therefore correctly classified as multicomponent. This can be seen from the last two cutouts.

## 7 RADIO SOURCE COUNTS

Given the flux density range $S$ to $S + \mathrm{d}S$, the observed number of sources d$N$ per unit area of the sky Ω is given by

$$\frac{\mathrm{d}N}{\mathrm{d}S_\nu} = \frac{n_\nu}{\Omega\Delta} \quad , \tag{13}$$

where $n_\nu$ is the total number of sources per unit area of the sky at flux density $S_\nu$ of bin width Δ in a survey area Ω. The observed raw source counts $n_\nu$ will decrease at faint flux densities due to incompleteness from varying sensitivity across the image. The non-detection of large sources due to resolution bias and the fraction of spurious (false) sources are also factors affecting the completeness of our radio catalogue. Therefore, in order to calculate the intrinsic source counts distribution we must apply these appropriate corrections to account for underestimation in the raw source counts. The final corrected observed number count is then given by:

$$\frac{\mathrm{d}N}{\mathrm{d}S_\nu} = \left[\frac{1}{\Omega\Delta}\right] \frac{f_{\mathrm{rb}}(1 - f_{\mathrm{ss}})n_\nu}{f_{\mathrm{nb}}} \ . \tag{14}$$

We show in Fig. 16 the observed (raw) and corrected Euclidean-normalized 1.28 GHz radio source counts derived for A2631 and ZwCL2341 at 5$\sigma$ threshold. The vertical error bars are the Poisson errors in the counts $\sqrt{n}$ (N. Gehrels 1986). We show details of the source count calculation in Table A1 along with the correction factors of noise bias $f_{\mathrm{nb}}$, resolution bias $f_{\mathrm{rb}}$. The fraction of false sources $f_{\mathrm{ss}}$ in Section 6.4 was found to be negligible and we therefore ignore its contribution to the incompleteness in the raw counts. We note that $f_{\mathrm{nb}}$ mainly a result of varying RMS due to primary beam correction and residual calibration errors (see Fig. 6) contributes mostly to the incompleteness especially at the faint end with correction factors $>> 10$. The other correction factors of $f_{\mathrm{rb}}$ are much smaller $<2$.

For comparison, we have overlayed the widely used (G. De Zotti et al. 2010) 1.4 GHz counts compilation along with other MeerKAT 1.28 GHz counts in the form of DEEP2 (T. Mauch et al. 2020), XMM Newton Large-Scale Structure (LSS) (C. Hale et al. 2023) and COSMOS (C. Hale et al. 2023). For the source count models we have used the older M. Bondi et al. (2008) together with newer models of C. Mancuso et al. (2017), SEMPER (M. Giulietti et al. 2025) and T-RECS (A. Bonaldi et al. 2023) for both AGN and SFG populations. We have scaled our counts to 1.4 GHz using a spectral index $\alpha = -0.7$ (V. Smolčić et al. 2017).

We see that the Euclidean normalized differential counts of the two clusters follow the trend of G. De Zotti et al. (2010) and the models for >1 mJy. We also see the well-known flattening at the sub-mJy level $0.1 < S_{\mathrm{T}} < 1$ mJy that indicates the emergence of a new population of SFGs dominating over the AGNs in both clusters. However, at this regime we see our counts are on average slightly higher than those of the other MeerKAT data and the models (see Section 7.2 for discussion). This pronounced 'bump' feature is more apparent in A2631 than ZwCl2341.

Our differential counts then drop off sharply at the faintest fluxes $S < 0.1$ mJy because they fall below the detection limits (where the flux densities are not reliable and still suffer from incompleteness effects) of our data (blue and red vertical dashed lines of A2631 and ZwCL2341, respectively). For the other MeerKAT 1.28 GHz deep counts of DEEP2, XMM LSS, and COSMOS, the incompleteness effects are negligible for $S < 0.1$ mJy due to the surveys deeper detection limits. This implies that their counts do not show such a rapid drop. This behaviour is consistent with the newer models of C. Mancuso et al. (2017), M. Giulietti et al. (2025), and A. Bonaldi et al. (2023).

### 7.1 Impact of cosmic variance on the source counts

Field-to-field variation in the number counts of radio sources are a direct consequence of the non-uniform large scale structure of the universe (P. Coles & L.-Y. Chiang 2000). This effect is called cosmic variance. I. Heywood, M. J. Jarvis & J. J. Condon (2013) showed that cosmic variance depends on a combination of area coverage and depth. For a given sample variance, the deeper the survey, the larger the volume sampled and therefore the smaller the area coverage required. The work of E. Retana-Montenegro et al. (2018) also showed that even for the same field split into smaller partitions the variation in the source counts (due to cosmic variance) is larger than the Poisson errors implying that cosmic variance could represent an important source of uncertainty in the source counts.

Since the majority of observations carried at 1.4 GHz are from single pointing observations of a few deg$^2$ at moderate (sub mJy) depth, the large spread observed in their source counts at <1 mJy is usually attributed to this cosmic variance effect (I. Prandoni et al. 2018). Therefore the differences we see in the source counts at the sub-mJy level of our data with the other MeerKAT data could be the cosmic variance effect in action.

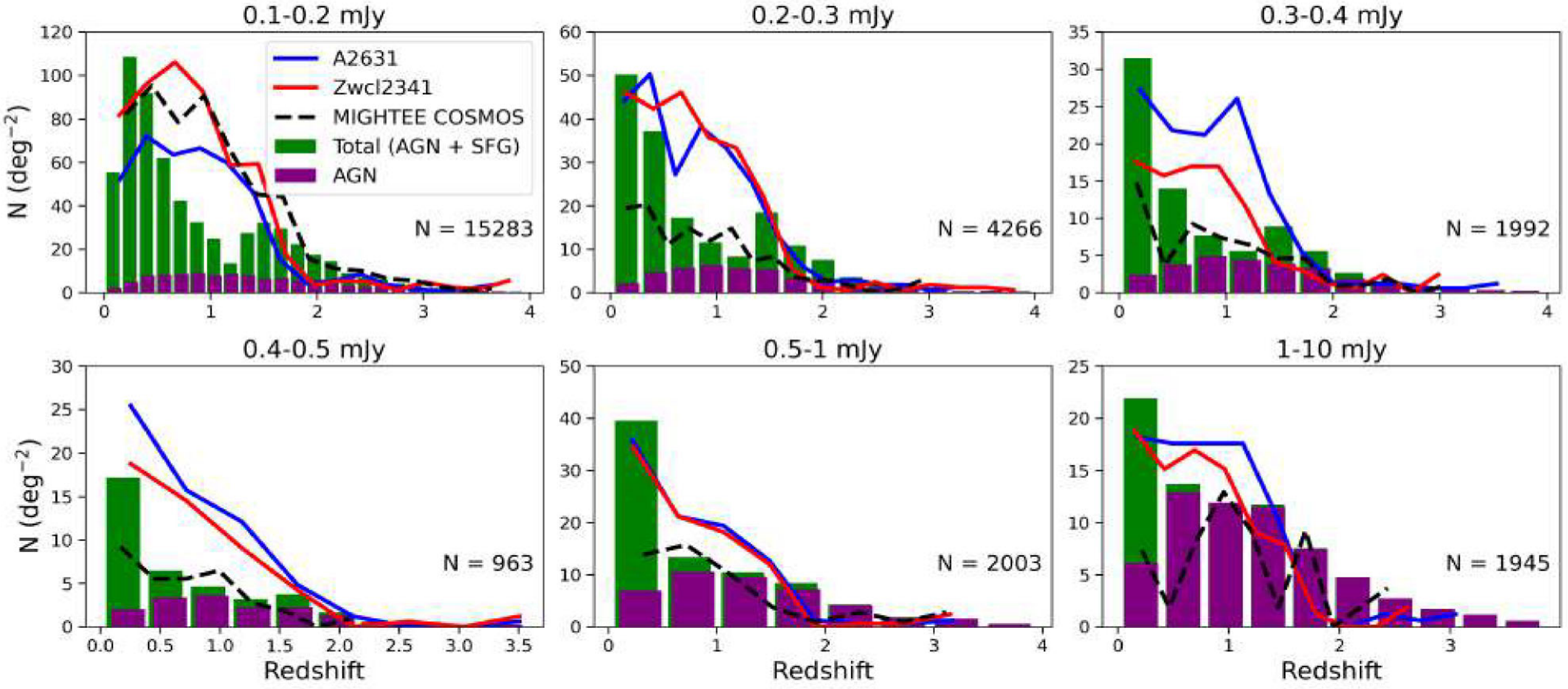

**Figure 19.** Redshift distribution of A. Bonaldi et al. (2023) (T-RECS) simulated catalogue with the radio sources of A2631 and ZwCL2341 at different flux density intervals. The redshifts of our radio catalogues were obtained through cross-match with HSC WIDE data. Photometric redshifts obtained from HSC WIDE are most accurate in the range $0.2 < z < 1.5$. The histograms indicate the AGN and Total (AGN + SFG) distribution of sources of the A. Bonaldi et al. (2023) simulated catalogue. The solid lines show the redshift distribution for the catalogues of A2631 and ZwCL2341, respectively. We have also shown the redshift distribution of MIGHTEE COSMOS data (dashed curve) obtained from the multiwavelength analysis performed in I. H. Whittam et al. (2022, 2023). The $y$-axis represents the source density for each catalogue in order for a proper comparison. Each panel represents a different flux density bin, increasing from left to right top to bottom. Notice the AGN population dominating with increase in flux density. The flux density range in the first five panels 0.1–1mJy are the region where our source counts are slightly higher compared to the A. Bonaldi et al. (2023) simulated model and the MIGHTEE MeerKAT data (C. Hale et al. 2023) (see Fig. 16).

### 7.2 Qualitative comparison with simulated models

We can compare the redshift distribution of the galaxy population at different flux density intervals of our data with that of simulated models to probe the cause of deviation of our source counts with these simulated models.

We show in Fig. 19 the HSC WIDE photometric redshift distribution resulting from the cross-match (see Section B) with the radio sources in A2631 and ZwCl2341 (blue and red solid lines, respectively) together with the redshift distribution of A. Bonaldi et al. (2023) (histograms) at different flux density intervals (displayed at the top of each panel) for $0 < z_{\rm phot} < 4$. We splitted the A. Bonaldi et al. (2023) simulated data into two groups of AGN (purple bars) and Total = AGN + SFG (green bars). We decrease bin size with increasing flux density as source statistics becomes increasingly sparse. The total number of sources from each simulated catalogue for each respective flux density interval is displayed within each plot. We note that photometric redshifts from MIZUKI are most robust in the range $0.2 < z_{\rm phot} < 1.5$ (M. Tanaka et al. 2018). We also plot the redshift distribution of MIGHTEE data for the central part of COSMOS (black dashed curves) from the radio/optical cross-matched catalogues of I. H. Whittam et al. (2022, 2023). The source counts of MIGHTEE COSMOS and A. Bonaldi et al. (2023) simulated data were both plotted in Fig. 16 (green triangles and orange solid curve, respectively) and hence are used as a comparison to the redshift distribution of our MeerKAT data.

The first five panels cover the flux density range $0.1 < S_{\rm 1.4GHz} < 1$ mJy and is where our observed radio source counts show a slight excess ('bump') as compared to the different evolutionary models and other MeerKAT data (including MIGHTEE COSMOS) in the source counts plot of Fig. 16. We can clearly see that in these panels there is a clear excess of sources in our MeerKAT data as compared to the A. Bonaldi et al. (2023) simulated sources and MIGHTEE COSMOS sources in agreement with the slightly higher counts observed in our MeerKAT data at the sub-mJy level. This excess is most notable for $z < 1.5$. The A. Bonaldi et al. (2023) simulated data are however in good agreement with MIGHTEE COSMOS, consistent with its agreement in the source counts plot of Fig. 16. As we reach flux density >1mJy (last panel), we see that the redshift distribution for our radio sources are in better agreement with the A. Bonaldi et al. (2023) model and MIGHTEE COSMOS with little excess of sources observed. The slightly higher counts observed in both fields at the sub mJy level could therefore be due to an excess of intermediate redshift $z < 1.5$ SF galaxies and/or AGN that exist in our cluster fields. S. Mandal et al. (2021) also observed a pronounced 'bump' in the source counts of the Elais-N1 field and found a similar result of source excess in flux density bins corresponding to the 'bump'.

## 8 SUMMARY AND CONCLUSIONS

We have presented deep MeerKAT radio catalogues at 1.28 GHz of two massive galaxy clusters A2631 and ZwCL2341 residing at the core of the *Saraswati* supercluster. Our catalogues contain 1999 and 2611 sources for A2631 and ZwCL2341, respectively, above a $5\sigma$ detection over an area of $\sim$1.6 deg$^2$. From these catalogues we have investigated systematic effects of astrometry and flux density scale accuracy. We derived the spectral index distribution between 998 and 1569 MHz and found it consistent with the value of $\sim -0.7$ often quoted in literature.

We also observed how the spectra of our radio sources changes with flux density. We found that by using cuts in the radio luminosity to define potential SFGs (RQ sources) and AGNs (RL sources), that these two classes can be distinguished based on their spectra. RQ sources are characterized by an overall flat spectra ($-0.2 < \alpha < -0.5$) and RL sources characterized by an overall steep spectra ($-0.5 < \alpha < -1$), with the differences largest at the extreme ends in flux density. Lastly, we derived the radio source counts including three correction factors taking into account noise bias, resolution bias and Eddington bias. We found that the source counts are generally in agreement with those derived from other surveys at 1.4 GHz for >1mJy.

For <1mJy, we observe the change in slope of the source counts ('transition region') indicating the emerging SFG population. Our counts are however slightly higher than those of new models and other deeper MeerKAT data at this transition region giving rise to a 'bump' feature that has been seen in other works (e.g. S. Mandal et al. 2021 for LOFAR). The fraction of false and multicomponent sources were investigated and they turned out to be negligible in both catalogues, thus ruling them out as potential causes to the inconsistencies of our counts with other data at the <1mJy level. From the qualitative comparison with the redshift distributions of A. Bonaldi et al. (2023) simulated evolutionary model and central region of COSMOS from MIGHTEE (I. H. Whittam et al. 2022, 2023) we conclude that the reason for the bump we observe is because the radio population in our data contains an excess of SFG and/or faint AGN at intermediate redshifts $z < 1.5$ which could be cosmic variance in play.

The follow up paper *MOSS 3* will combine these radio catalogs with the wealth of multiwavelength public data available (optical, infrared, etc) including spectroscopic redshifts and source classifications. This will enable for a more detailed study of the radio spectral properties of AGN and star-forming galaxies at various positions within the cluster core region in an attempt to understand the effect cluster environments have on the radio galaxy population.

## ACKNOWLEDGEMENTS

This work was supported by the Swiss National Science Foundation (SNSF) under funding reference 200021_213076 'Galaxy evolution in the cosmic web'. We acknowledge the funding contribution from the Swiss government excellence scholarship (Ref:2022.0300) and SS acknowledges the support from the Estonian Research Council grant PSG1045. The financial assistance offered by the South African Radio Astronomy Observatory (SARAO) towards this research is also acknowledged. This work was also supported by the European Regional Development Fund under grant agreement PK.1.1.10.0007 (DATACROSS). Lastly, we thank the referee for their insightful comments that lead to an overall improvement in the paper.

## DATA AVAILABILITY

The corrected and uncorrected radio catalogues used in this work are provided as supplementary data. The structure of these catalogues are shown in Tables 3 and 4 for A2631 and ZwCl2341, respectively. The code used to generate the figures in this work are available on Github (https://github.com/Kincaidr/Saraswati.git).

## APPENDIX A: APPENDIX A: SOURCE COUNTS

**Table A1.** Radio source counts scaled to 1.4 GHz of the inner 1.6 square deg region of A2631 (top) and ZwCL2341 (bottom) normalized to Euclidean geometry.

| $S_I$ (mJy) | $\Delta S$ (mJy) | $N$ | $S^{5/2}n(S)$ ($Jy^{1.5}$ $sr^{-1}$) | Error ($Jy^{1.5}$ $sr^{-1}$) | $fnb$ | $frb$ |
|---|---|---|---|---|---|---|
| (1) | (2) | (3) | (4) | (5) | (6) | (7) |
| 0.0399 | 0.0138 | 1 | 0.001 45 | 0.001 45 | 142.31 | 1.64 |
| 0.0565 | 0.0195 | 30 | 0.0735 | 0.0134 | 25.37 | 1.72 |
| 0.0800 | 0.0276 | 131 | 0.5410 | 0.0473 | 5.99 | 1.48 |
| 0.1134 | 0.0391 | 296 | 2.061 | 0.120 | 2.40 | 1.31 |
| 0.1606 | 0.0554 | 444 | 5.214 | 0.247 | 1.53 | 1.18 |
| 0.2276 | 0.0785 | 374 | 7.407 | 0.383 | 1.22 | 1.10 |
| 0.3224 | 0.1112 | 270 | 9.018 | 0.549 | 1.10 | 1.05 |
| 0.4568 | 0.1576 | 133 | 7.491 | 0.650 | 1.04 | 1.02 |
| 0.6472 | 0.2232 | 94 | 8.928 | 0.921 | 1.01 | 1.01 |
| 0.9170 | 0.3163 | 55 | 8.810 | 1.19 | 1.00 | 1.00 |
| 1.2992 | 0.4481 | 51 | 13.776 | 1.93 | 1.00 | 1.00 |
| 1.8406 | 0.6349 | 21 | 9.566 | 2.09 | 1.00 | 1.00 |
| 2.6078 | 0.8995 | 27 | 20.741 | 3.99 | 1.00 | 1.00 |
| 3.6947 | 1.2743 | 11 | 14.250 | 4.30 | 1.00 | 1.00 |
| 5.2346 | 1.8055 | 11 | 24.031 | 7.25 | 1.00 | 1.00 |
| 7.4163 | 2.5580 | 12 | 44.210 | 12.8 | 1.00 | 1.00 |
| 10.5074 | 3.6241 | 11 | 68.342 | 20.6 | 1.00 | 1.00 |
| 14.8867 | 5.1346 | 7 | 73.342 | 27.7 | 1.00 | 1.00 |
| 21.0914 | 7.2746 | 4 | 70.676 | 35.3 | 1.01 | 1.00 |
| 29.8820 | 10.3066 | 9 | 268.171 | 89.4 | 1.01 | 1.00 |
| 42.3364 | 14.6023 | 2 | 100.498 | 71.1 | 1.02 | 1.00 |
| 59.9817 | 20.6883 | 1 | 84.739 | 84.7 | 1.03 | 1.00 |
| 84.9814 | 29.3110 | 0 | 0.0 | 0.0 | 1.04 | 1.00 |
| 120.4007 | 41.5275 | 1 | 240.989 | 240.989 | 1.06 | 1.00 |
| 170.5822 | 58.8356 | 1 | 406.401 | 406.401 | 1.09 | 1.00 |
| $S_I$ (mJy) | $\Delta S$ (mJy) | $N$ | $S^{5/2}n(S)$ ($Jy^{1.5}$ $sr^{-1}$) | Error ($Jy^{1.5}$ $sr^{-1}$) | $fnb$ | $frb$ |
| (1) | (2) | (3) | (4) | (5) | (6) | (7) |
| 0.0301 | 0.0107 | 20 | 0.0185 | 0.004 14 | 57.22 | 1.48 |
| 0.0432 | 0.0154 | 107 | 0.1698 | 0.0164 | 17.68 | 1.69 |
| 0.0619 | 0.0220 | 218 | 0.5936 | 0.0402 | 5.18 | 1.65 |
| 0.0887 | 0.0316 | 464 | 2.167 | 0.101 | 2.51 | 1.43 |
| 0.1272 | 0.0453 | 521 | 4.175 | 0.183 | 1.59 | 1.26 |
| 0.1822 | 0.0649 | 438 | 6.022 | 0.288 | 1.26 | 1.15 |
| 0.2612 | 0.0930 | 310 | 7.312 | 0.415 | 1.11 | 1.08 |
| 0.3743 | 0.1332 | 160 | 6.475 | 0.512 | 1.04 | 1.03 |
| 0.5364 | 0.1910 | 113 | 7.845 | 0.738 | 1.02 | 1.01 |
| 0.7687 | 0.2737 | 58 | 6.908 | 0.907 | 1.01 | 1.00 |
| 1.1016 | 0.3922 | 42 | 8.582 | 1.32 | 1.00 | 1.00 |
| 1.5787 | 0.5620 | 35 | 12.27 | 2.07 | 1.01 | 1.00 |
| 2.2624 | 0.8054 | 29 | 17.44 | 3.24 | 1.00 | 1.00 |
| 3.2422 | 1.1543 | 21 | 21.67 | 4.73 | 1.00 | 1.00 |
| 4.6465 | 1.6542 | 17 | 30.09 | 7.30 | 1.00 | 1.00 |
| 6.6588 | 2.3706 | 13 | 39.48 | 10.95 | 1.00 | 1.00 |
| 9.5428 | 3.3973 | 7 | 36.47 | 13.8 | 1.01 | 1.00 |
| 13.6757 | 4.8686 | 9 | 80.44 | 26.8 | 1.01 | 1.00 |
| 19.5987 | 6.9772 | 8 | 122.7 | 43.4 | 1.02 | 1.00 |
| 28.0868 | 9.9991 | 5 | 131.5 | 58.8 | 1.03 | 1.00 |
| 40.2512 | 14.3297 | 3 | 135.4 | 78.2 | 1.05 | 1.00 |
| 57.6840 | 20.5358 | 2 | 154.9 | 109.5 | 1.07 | 1.00 |
| 82.6668 | 29.4299 | 2 | 265.7 | 187.9 | 1.11 | 1.00 |
| 118.4697 | 42.1759 | 2 | 455.8 | 322.3 | 1.18 | 1.00 |
| 169.7788 | 60.4423 | 0 | 0.0 | 0.0 | 1.28 | 1.00 |

*Note.* (1) Bin centre, (2) bin width, (3) number of sources in bin, (4, 5) euclidean-normalized counts and error, (6) noise bias, and (7) resolution bias correction.

## APPENDIX B: HSC/MEERKAT CROSS-MATCH AND VLA/MEERKAT FLUX DENSITY COMPARISON

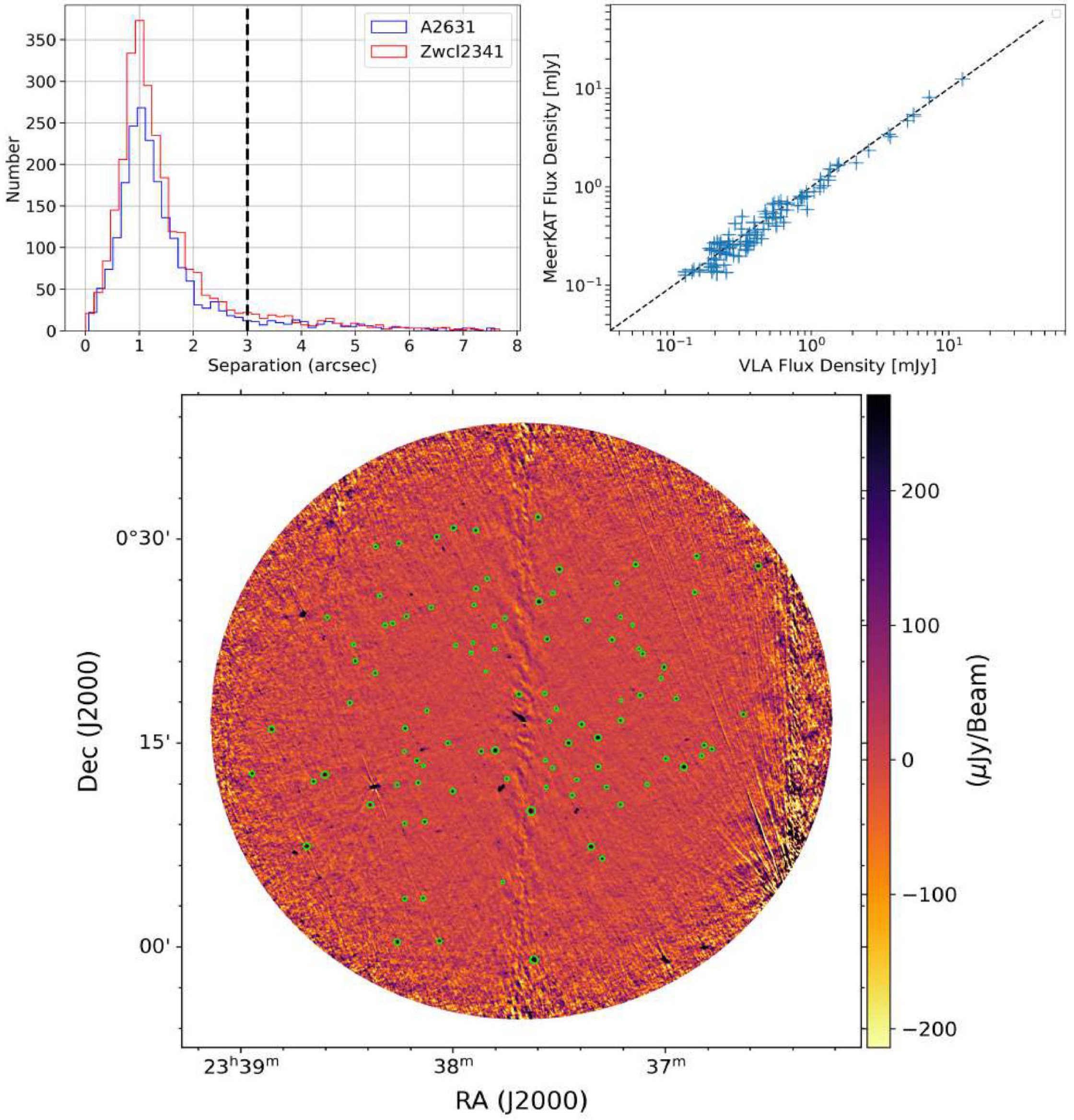

**Figure B1.** Top left: MeerKAT/HSC WIDE distance distribution after matching to nearest neighbour. The dashed vertical black line indicates the radius in which we chose to perform the MeerKAT/HSC WIDE cross-match to obtain the photometric redshifts from the MIZUKI table (See Section 7.2). It can be seen that matching beyond this radius will result in spurious sources. 95 per cent of our MeerKAT sources have an HSC WIDE counterpart after cross-match at this radius. Top right: the MeerKAT/EVLA flux density comparison for 100 manually selected sources. The vertical and horizontal lines indicate the reported error associated with the total flux density measurement of CASA IMFIT from region files on the MeerKAT and EVLA image, respectively. Black dotted line indicates the 1:1 curve. Bottom: high resolution (0.6 arcsec) deep EVLA 1.6 GHz image of the inner (0.4 deg$^2$) region of A2631 used for the EVLA/MeerKAT flux density consistency check. We convolved the higher resolution EVLA image with the beam of MeerKAT (8 arcsec) to match the beam sizes. The green circles on the image are the region files of 100 manually selected bright sources used to compute the total flux density on both images. The EVLA/MeerKAT total flux density show a tight correlation that decrease in scatter with increasing flux density.

This paper has been typeset from a TEX/LATEX file prepared by the author.